\documentclass[prl,twocolumn,longbibliography,superscriptaddress,preprintnumbers]{revtex4-2}
\usepackage{comment}
\usepackage{bm}
\usepackage{dsfont}
\usepackage{graphicx}
\usepackage{physics}
\usepackage{epsfig}
\usepackage{amsfonts}
\usepackage{epstopdf}
\usepackage{balance}
\usepackage[dvipsnames]{xcolor}
\usepackage{calc}
\usepackage{natbib}
\usepackage[colorlinks,
            linkcolor=blue,
            anchorcolor=blue,
            citecolor=blue,
            urlcolor=blue]{hyperref}
\usepackage{lipsum}
\usepackage[version=3]{mhchem} 
\usepackage{color}
\usepackage{soul}
\usepackage[etex=true,export]{adjustbox}
\usepackage{makecell}
\usepackage{multirow}
\usepackage{tabularx,multirow,array,diagbox}
\usepackage{adjustbox}
\usepackage{booktabs}
\makeatletter
\renewcommand{\maketag@@@}[1]
{\hbox{\m@th\normalsize\normalfont#1}}%

\newcommand\bl[1]{\boldsymbol{#1}}
\newcommand\wt[1]{\widetilde{#1}}

\makeatother
  
\begin{document}
\title{Quantum-Geometric Design of Lattice Generalized Landau Levels}

\author{Bohao Li}
%\email{bohaoli@whu.edu.cn}
\affiliation{School of Physics and Technology, Wuhan University, Wuhan 430072, China}
\author{Fengcheng Wu}
\email{wufcheng@whu.edu.cn}
\affiliation{School of Physics and Technology, Wuhan University, Wuhan 430072, China}
\affiliation{Wuhan Institute of Quantum Technology, Wuhan 430206, China}

\begin{abstract}
We design lattice models with tailored quantum geometry, including generalized Landau levels (LLs) satisfying the integrated trace condition and higher-Chern bands with ideal quantum geometry. Our models with $N=2$, $3$, and $4$ sublattices include a generalized Haldane model ($N=2$ honeycomb lattice model) with Gaussian-decaying hoppings realizable in twisted bilayer MoTe$_2$, and $N \geq 3$ models with exponentially decaying hoppings. Exact diagonalization reveals fractional Chern insulators in the generalized zeroth LL bands of all three models, a Moore–Read state in the generalized first LL band of the $N=4$ model, and various interaction-driven topological phases—including integer and fractional anomalous Hall crystals and a multicomponent Halperin state—in the ideal higher-Chern band of the $N=3$ model. Informed by quantum geometry, our work provides a pathway for lattice realizations of Landau-level and beyond-Landau-level physics.
\end{abstract}
\maketitle

\textit{Introduction.—}
Fractional quantum Hall insulators (FQHIs) arise in LLs of two-dimensional electrons under strong magnetic fields, hosting quasiparticles with fractional statistics \cite{Tsui1982Two,Laughlin1983}. Their lattice analogs—fractional Chern insulators (FCIs)—replace LLs with topological Chern bands \cite{Tang2011,Sun2011,Neupert2011,Regnault2011Fractional,Sheng2011Fractional}. Recent experiments have realized FCIs at zero magnetic field in moir\'e materials, including twisted MoTe$_2$ (tMoTe$_2$)\cite{Cai2023,Zeng2023,Park2023,Xu2023Observation} as well as multilayer rhombohedral graphene aligned with hBN \cite{Lu2024Fractional,Xie2025Tunable}. These developments motivate the examination of a central question: how to systematically design lattice bands that capture the essential physics of LLs while preserving lattice translation symmetry.

A defining property of LLs is their unique quantum geometry, characterized by the momentum-independent quantum metric and Berry curvature that satisfy an index-dependent trace condition \cite{Roy2014Band}. However, LL wave functions are constrained by magnetic translational symmetry (MTS), which is incompatible with ordinary lattice translations. Consequently, lattice representations of LLs are generally quasiperiodic, as exemplified by the Kapit–Mueller model \cite{Kapit2010Exact}. While it realizes an exact flat band as the lattice analog of the zeroth LL, it inherits MTS, a feature shared by extensions to non-Bravais lattices \cite{Xu2020Building,Dong2020Exact}, higher LLs \cite{Atakisi2013Landau,Shen2026exact}, and higher-Chern number states \cite{Behrmann2016Model}.

In this Letter, we use quantum geometry as a guiding criterion to design lattice models hosting generalized LLs and higher-Chern bands. Generalized LLs, obtained from spatially modulated LL wave functions in continuum space, retain the essential geometric properties of LLs \cite{liu2025theoryofgLL,li2025Variational,li2026abelian}. By constructing lattice Bloch states from these wave functions, we realize Chern bands that inherit LL-like quantum geometry while respecting lattice translation symmetry, establishing a direct and symmetry-compatible connection between LL physics and lattice systems. This construction also produces higher-Chern bands with ideal quantum geometry, which are of great theoretical \cite{Ledwith2022IdealChern,Wang2022Hierarchy} and experimental \cite{dong2025observation,li2026FractionalHigh} interest as well.

We construct explicit lattice models with sublattice number $N=2,3$, and $4$, realizing generalized LLs and ideal higher-Chern bands. The $N=2$ model on the honeycomb lattice can be viewed as a generalized Haldane model with Gaussian-decaying hopping parameters, and can be quantitatively realized in tMoTe$_2$ at the magic angle. For $N \geq 3$ models, requiring all bands to be exactly flat yields exponentially-decaying hoppings.  Exact diagonalization (ED) identifies Abelian FCIs at filling factors $\nu=1/3$ and $2/3$ in the generalized 0LL bands across all three models, and a non-Abelian Moore–Read state at $\nu=1/2$ in the generalized 1LL band in the $N=4$ model. In higher-Chern bands, we further obtain both integer and fractional anomalous Hall crystal states as well as multicomponent Halperin states. These results demonstrate that a rich variety of correlated topological phases can be realized within a single, unified family of lattice models with engineered quantum geometry.

\begin{figure}[t]
    \includegraphics[width=1.\columnwidth]{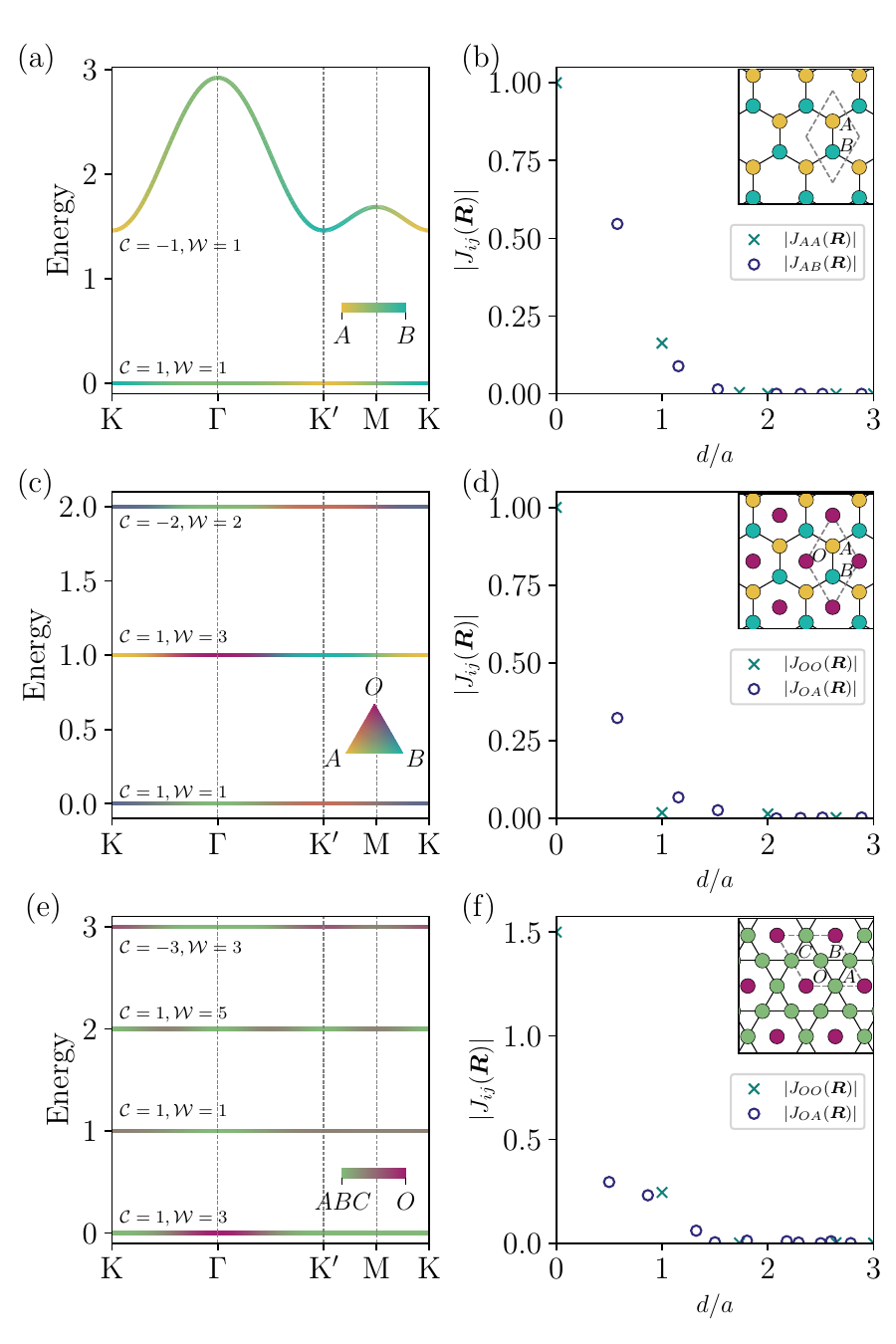}
    \caption{Band structures (left panels) and hopping amplitudes (right panels) for the $N=2$, $3$, and $4$ models. Left panels: Colors encode the sublattice weights. Right panels: The corresponding hopping amplitudes as a function of distance $d$, measured in unit of the lattice constant $a$, are shown in (b), (d), and (f). Insets display the lattice structures, with dashed lines indicating the unit cells.} 
    \label{fig:1}
\end{figure}

\textit{Generalized LLs on lattice.—} 
We start by reviewing the construction of generalized LLs in the continuum space. To characterize the states, we analyze the quantum geometric tensor $ \mathcal{Q}_{\bl k}$, decomposed into the quantum metric $g_{\bl{k}}$ and Berry curvature $\Omega_{\bl k}$ through $( \mathcal{Q}_{\bl k})_{ab} = (g_{\bl k})_{ab} + \frac{i}{2}\epsilon_{ab}\Omega_{\bl k}$, with $\epsilon_{ab}$ the Levi-Civita symbol. Their Brillouin-zone integrals give the quantum weight $\mathcal{W} = \frac{1}{2\pi} \int d\bl{k} \mathrm{Tr}[g_{\bl{k}}]$ and Chern number $\mathcal{C} = \frac{1}{2\pi} \int d\bl{k} \Omega_{\bl{k}}$ \cite{Onishi2025Quantum}.

For a $\mathcal{C}=1$ Chern band satisfying the ideal quantum geometry condition $\mathrm{Tr}[g_{\bl k}]=\Omega_{\bl k}$, the Bloch state takes the form of the generalized zeroth LL (0LL) state \cite{Wang2021Exact,Tarnopolsky2019,Ledwith2020,Wang2021Chiral},
\begin{equation}
\label{g0LL}
\begin{aligned}
\Theta_{0,\bl k}(\bl r)=\wt{\mathcal{N}}_{0, \bl k}\mathcal B(\bl r)\Psi_{0,\bl k}(\bl r)
\end{aligned}     
\end{equation}
where $\mathcal{B}(\bl r)$ is a spatial modulation function, $\Psi_{n,\bl k}(\bl r)$ is the magnetic Bloch wave function in the symmetric gauge for the $n$th LL ($n$LL) \cite{Haldane2018modular}, and $\wt{\mathcal{N}}_{0, \bl k}$ is the normalization factor. For $\Theta_{0,\bl k}(\bl r)$ to satisfy the Bloch translational symmetry, $\mathcal B(\bl r)$ and $\Psi_{0,\bl k}(\bl r)$ should obey opposite MTS,
\begin{equation}
\begin{aligned}
\Psi_{0,\bl k}(\bl r + \bl R_i ) =&-e^{+i \frac{1}{2\ell^2} \bl R_i \times \bl r } e^{i \bl k \cdot \bl R_i}\Psi_{0,\bl k}(\bl r )   \\
\mathcal B(\bl r + \bl R_i) =& -e^{-i \frac{1}{2\ell^2} \bl R_i \times \bl r }\mathcal B(\bl r),
\end{aligned}
\label{eq:MTS}
\end{equation}
where $\bl R_{1}$ and $\bl R_{2}$ are the basis vectors of a unit cell. The unit cell for the Bloch state $\Theta_{0,\bl k}(\bl r)$ coincides with the magnetic unit cell of the LL state $\Psi_{0,\bl k}(\bl r)$, which has an area of $\mathcal{A}_0=2\pi \ell^2$.   $\Theta_{0,\bl k}(\bl r)$ has been further extended to generalized $n$th LL wave function $\Theta_{n,\bl k}(\bl r)$ by applying Gram–Schmidt orthogonalization to a set of density-modulated basis functions $e_{n,\bl k}(\bl r)$ \cite{liu2025theoryofgLL}, defined as
\begin{equation}
\label{enLL}
\begin{aligned}
e_{n,\bl k}(\bl r)=\mathcal B(\bl r)\Psi_{n,\bl k}(\bl r).
\end{aligned}     
\end{equation}
The quantum geometry of $\Theta_{n,\bl k}(\bl r)$ becomes momentum dependent, but satisfies the integrated form of the trace condition $\mathcal{W}=(2n+1) \mathcal C=2n+1$. 
%Notably, this state has recently been adopted as a variational ansatz for the Bloch state, with a high overlap between the two serving as a key indicator for diagnosing both Abelian and non-Abelian fractionalized states \cite{li2026abelian}.

We now construct a lattice realization of the generalized LLs, extending their original continuum formulation. In an $N$-sublattice system, we introduce $\ket{\Phi_{n,\bl k}}$ for $0\le n\le N-2$ as the lattice version of the generalized $n$LL state. To this end, we first construct the density-modulated basis $\ket{e_{n,\bl k}}$ by sampling the functions $e_{n,\bl k}(\bl r)$ on lattice,
\begin{equation}
\begin{small}
\begin{aligned}
\ket{e_{n,\bl k}} = \frac{1}{\sqrt{N_{d}}}\sum_{\bl R}\sum_{i=1}^Ne_{n,\bl k}(\bl R+\bl \tau_i)\ket{\bl R+\bl \tau_i},
\end{aligned}
\end{small}
\label{enk}
\end{equation}
where $N_{d}$ is the number of unit cells, $\bl R$ is the lattice vector, $\bl\tau_i$ is the  position of the $i$th sublattice within the unit cell, and   $\ket{\bl R+\bl \tau_i}$ is the corresponding orbital. $\ket{e_{n,\bl{k}}}$ is determined, via translational properties given in Eq.~\eqref{eq:MTS}, by the values of $\mathcal{B}(\bl\tau_i)$ for $i=1,...,N$. Assuming the states $\ket{e_{n,\bl k}}$ are linearly independent for $0\le n\le N-2$ (see Supplemental Metarial (SM) \cite{SM} for more discussion), $\ket{\Phi_{n,\bl k}}$ are obtained by applying Gram–Schmidt orthogonalization to $\ket{e_{n,\bl k}}$,
\begin{equation}
\begin{small}
\begin{aligned}
\ket{\Phi_{n,\bl k}}\!=\!
\begin{cases}
\mathcal N_{0,\bl k}\!\ket{e_{0,\bl k}}\!\!&\!\! n = 0,\\
\mathcal{N}_{n,\bl k}\Bigl[\ket{e_{n,\bl k}}\!-\!\!\textstyle\sum\limits_{m=0}^{n-1}\!\bra{\Phi_{m,\bl k}}\!e_{n,\bl k}\rangle\!\ket{\Phi_{m,\bl k}}\Bigr]\!\!&\!\!1\!\le n\!\le N-2,
\end{cases}
\end{aligned}
\end{small}
\end{equation} 
where $\mathcal{N}_{n,\bl k}$ is the normalization factor.
Here $\ket{\Phi_{n,\bl k}}$ inherits the quantum geometric properties of the generalized $n$LL, carries Chern number $\mathcal{C}=1$ and satisfies the integrated form of the trace condition $\mathcal{W}=(2n+1) \mathcal C=2n+1$. Specifically, $\ket{\Phi_{0,\bl k}}$ has ideal quantum geometry.

The remaining Bloch state $\ket{\Phi_{N-1,\bl k}}$, however, is qualitatively different. It is determined by the completeness relation $\sum_{n=0}^{N-1}\ket{\Phi_{n,\bl{k}}}\bra{\Phi_{n,\bl{k}}}=\mathds{1}$ on the lattice and carries Chern number $\mathcal{C}=-(N-1)$.  Remarkably, $\ket{\Phi_{N-1,\bl{k}}}$ is anti-holomorphic in $\bl{k}$ up to a normalization factor (see SM \cite{SM} for a proof), thereby satisfying the trace condition $\mathrm{Tr}[g_{\bl k}] = |\Omega_{\bl k}|$ and yielding a quantum weight of $\mathcal{W}=N-1$. Therefore, $\ket{\Phi_{N-1,\bl k}}$ realizes an ideal (higher) Chern band for $N=2$ ($N \geq 3$).

\begin{figure}[t]
    \includegraphics[width=1.\columnwidth]{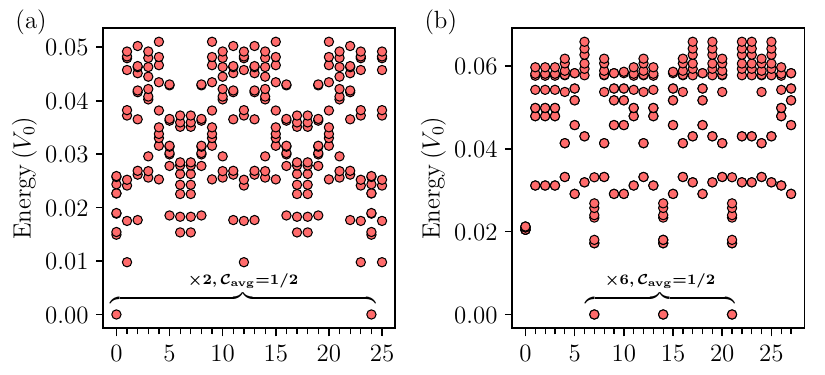}
    \caption{(a,b) ED spectra at $\nu=1/2$ for the generalized 1LL band in the $N=4$ model on clusters with $N_s=26$ and $28$, respectively. Energy is measured relative to the ground state.} 
    \label{fig:1LL}
\end{figure}

\textit{Models.—}
We formulate a Hamiltonian with $\ket{\Phi_{n,\bl k}}$ as its eigenstate,
\begin{equation}
\begin{small}
\begin{aligned}
\hat{H} = & \sum_{\bl k} \sum_{n=0}^{N-1} \mathcal{E}_{n,\bl k} \,
\ket{\Phi_{n,\bl k}}\bra{\Phi_{n,\bl k}}\\
= & \sum_{\bl R,\bl R'}\sum_{i,j} J_{ij}(\bl R-\bl R') \,
\ket{\bl R+\bl\tau_i}\bra{\bl R'+\bl\tau_j}.
\end{aligned}
\end{small}
\end{equation} 
Here $\mathcal{E}_{n,\bl k}$ is the band energy and $J_{ij}(\bl R-\bl R')$ are the resulting real-space hopping parameters, which respect lattice translational symmetry.

The $N=2$ model is defined on a honeycomb lattice illustrated in the inset of Fig.~\ref{fig:1}(b). The unit cell contains two sublattices A and B located at $\bl{\tau}_{1,2} = \frac{a}{\sqrt{3}}\left[\frac{\sqrt{3}}{2}, \pm \frac{1}{2}\right]$, where $a$ is the lattice constant.
We set the magnitude of $\mathcal{B}(\bl{\tau}_i)$ to unity throughout this work, while its phase corresponds to a gauge choice. By choosing $\mathcal{E}_{0,\bl k}=0$ and $\mathcal{E}_{1,\bl k}=\mathcal{N}_{0,\bl k}^{-2}\mathcal A_0$, we obtain the band structure shown in Fig.~\ref{fig:1}(a). Both bands realize the generalized 0LL with ideal quantum geometry, consisting of a flat $\mathcal{C}=1$ band at zero energy and a dispersive $\mathcal{C}=-1$ band at higher energy. The color in Fig.~\ref{fig:1}(a) represents the sublattice decomposition, highlighting that the band topology arises from sublattice hybridization.
 
With the above choice of band energies, $J_{ij}(\bl R)$ can be evaluated analytically in the $N=2$ model~\cite{SM},
\begin{equation}
\label{eq:J_N=2}
\begin{aligned}
J_{ij}(\bl R)= & \eta_{\bl R}\,e^{i\bl u_{ij} \times \bl R/\ell^2}  e^{-d^2/(4\ell^{2})},
\end{aligned}
\end{equation}
where $d=|\bl R+\bl\tau_i-\bl\tau_j|$ is the hopping distance, $\bl u_{11}=\bl \tau_2$, $\bl u_{22}=\bl \tau_1$, and $\bl u_{12}=(\bl \tau_1+\bl \tau_2)/2$. The factor $\eta_{\bl R}$ is $(-1)^{m+n+mn}$ for $\bl R=m\bl R_1+n\bl R_2$, where $\bl R_{1}=a\left[1,0\right]$ and $\bl R_{2}=a\left[-\frac{1}{2},\frac{\sqrt{3}}{2}\right]$.  Here we set $\mathcal B(\bl \tau_1)=1$ and $\mathcal B(\bl \tau_2)=-e^{-\frac{i}{2\ell^2}\bl \tau_1\times\bl \tau_2}$ so that the nearest-neighbour hopping is real and positive. As shown in Fig.~\ref{fig:1}(b), $|J_{ij}(\bl R)|$ exhibits Gaussian decay with increasing hopping distance, compatible with the theorem in Ref.~\cite{Chen2014impossibility}. This model can be quantitatively realized in tMoTe$_2$ at the magic angle \cite{Wu2019Topological, Devakul2021}, where the two topmost moir\'e valence bands effectively form a honeycomb lattice model with hopping parameters closely matching those in Eq.~\eqref{eq:J_N=2} \cite{SM}. Upon truncating $J_{ij}(\bl R)$ to next-nearest-neighbor hopping, the $N=2$ model reduces to the Haldane model.

%We can further establish an exact mapping to the Kapit–Mueller model on the honeycomb lattice \cite{Kapit2010Exact,Dong2020Exact}. Specifically, under the site-dependent gauge transformation via the $\mathcal{B}(\bl r)$ function, the Kapit–Mueller model becomes identical to our $N=2$ model \cite{SM}.
  
We turn to the $N=3$ model defined on the honeycomb lattice augmented by an additional site O located at $\bl{\tau}_3 = [0, 0]$, as shown in the inset of Fig.~\ref{fig:1}(d). To mimic the equally spaced spectrum of conventional LLs, we choose  $\mathcal{E}_{0,\bl k}=0$, $\mathcal{E}_{1,\bl k}=1$, and $\mathcal{E}_{2,\bl k}=2$, yielding three flat bands with $(\mathcal{C}, \mathcal{K}) = (+1, 1)$, $(+1, 3)$, and $(-2, 2)$, as shown in Fig.~\ref{fig:1}(c). The lowest and middle bands realize the
generalized 0LL and 1LL on the lattice, respectively, while the topmost band is an ideal higher Chern band. Figure~\ref{fig:1}(d) shows that the hopping amplitudes decay exponentially at long distances \cite{SM}.

The $N=4$ model is defined on a kagome lattice formed by sublattices A, B, and C, augmented by an additional site O at the center of each hexagon [inset of Fig.~\ref{fig:1}(f)]. %The sublattices A, B, C, and O are located at $\bl\tau_1 = a[1/2, 0]$, $\bl\tau_{2,3} = a[\pm1/4, \sqrt{3}/4]$, and $\bl{\tau}_4 = [0, 0]$, respectively. 
We choose $\mathcal{E}_{0,\bl{k}}=1$, $\mathcal{E}_{1,\bl{k}}=0$, $\mathcal{E}_{2,\bl{k}}=2$, and $\mathcal{E}_{3,\bl{k}}=3$, deliberately placing the generalized 1LL at the lowest energy in contrast to the conventional LL ordering. This yields four flat bands, ascending in energy, with $(\mathcal{C}, \mathcal{K}) = (+1, 3)$, $(+1, 1)$, $(+1, 5)$, and $(-3, 3)$, as shown in Fig.~\ref{fig:1}(e). The lowest, second, and third bands realize the generalized 1LL, 0LL, and 2LL, respectively, while the top band is an ideal higher-Chern band. Despite the inverted LL ordering, the hopping amplitudes remain exponentially decaying at long distances \cite{SM}, as shown in Fig.~\ref{fig:1}(f). This construction highlights the enhanced versatility of lattice models, where the 1LL can be realized as the lowest-energy band. We note that a more uniform real-space lattice sampling, as $N$ increases, smooths the quantum geometry in momentum space.

\textit{Abelian FCIs.—}
We study interacting electrons by projecting the interaction onto a selected band.
\begin{equation}
\begin{aligned}
\label{H1}
\hat{\mathcal H}_n=\mathcal P_n\hat{\mathcal H}\mathcal P_n,\quad
\hat{\mathcal H} = \frac{1}{2}\sum_{\bl r,\bl r'} V(|\bl r-\bl r'|)\!:\!\hat n_{\bl r}\hat n_{\bl r'}\!:\;,
\end{aligned}
\end{equation}
where $\mathcal P_n$ projects onto the Bloch state $\ket{\Phi_{n,\bl k}}$, $\hat n_{\bl r}$ is the density operator at site $\bl r$, and the colons denote normal ordering with respect to the vacuum, defined by an empty $\ket{\Phi_{n,\bl k}}$ band. In $\hat{\mathcal H}$, $V(r)$ is taken to be either a Yukawa potential $U_0\frac{\exp(-\kappa r/a)}{r/a}$, or the nearest-neighbor interaction $V_0 \delta_{r, d_0}$ with $d_0$ the nearest-neighbor distance.

We first study many-body physics in the generalized 0LL band and perform ED for $\hat{\mathcal{H}}_0$ at fillings $\nu=1/3$ and $2/3$ across the three models. Here $\nu$ is defined as the number of electrons per unit cell in the selected band. For the Yukawa potential with $\kappa=0.25$, the exact diagonalization (ED) spectrum reveals a threefold quasi-degenerate ground state with a clear energy gap. The fractionally quantized many-body Chern number is  $\mathcal C_{\text{avg}}=1/3$ and $2/3$ (averaged over the ground state manifold), respectively, for $\nu=1/3$ and $2/3$. These results indicate the emergence of robust Abelian FCIs \cite{SM}.

\textit{Moore-Read state.—}
We then investigate the many-body physics of the generalized 1LL band in the $N=4$ model, where the variation of quantum geometry is suppressed compared to that in the generalized 1LL band of the $N=3$ model. We perform ED calculation in $\hat{\mathcal{H}}_1$ with nearest-neighbor interaction at $\nu=1/2$, using clusters of sizes $N_s=26$ and $28$, as shown in Fig.~\ref{fig:1LL}(a-b). The low-energy spectra exhibit characteristic ground-state degeneracies that depend on the parity of the electron number $N_e = N_s/2$: a sixfold quasi-degeneracy for even $N_e$ in the $N_s=28$ cluster, and a twofold quasi-degeneracy for odd $N_e$ in the $N_s=26$ cluster. This parity-dependent degeneracy is a hallmark of the Moore-Read (MR) state \cite{Read2000Paired}. The momentum sectors of these quasi-degenerate ground states are consistent with the (1,2) generalized Pauli principle for the MR state \cite{Haldane1991Fractional,Bernevig2012Emergent}. The topological nature of this phase is further confirmed by the many-body Chern number $\mathcal C_{\text{avg}}=1/2$.

To further probe the nature of the state, we examine the particle entanglement spectrum (PES) \cite{SM}. In both clusters, the PES displays an entanglement gap, separating the low-lying PES levels from the generic continuum. The counting of the levels below the gap matches the quasihole counting expected for the MR state \cite{Read2006Wavefunctions}.

\begin{figure}[t]
    \includegraphics[width=1.\columnwidth]{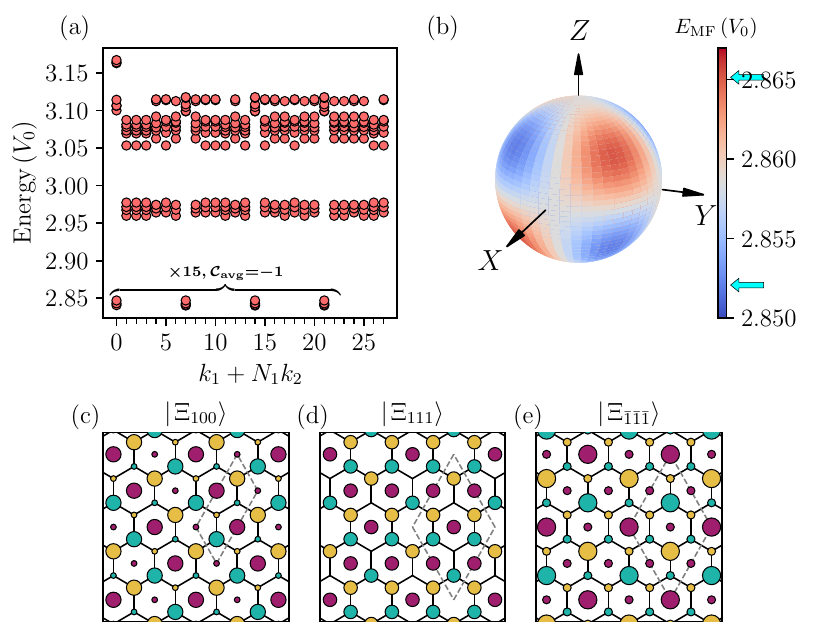}
    \caption{(a) ED spectrum at $\nu=1/2$ for the $\mathcal{C}=-2$ band in the $N=3$ model, calculated on a cluster with $N_s=28$. (b) Mean-field energy of Slater determinant states for the same system, parameterized on the Bloch sphere. Arrows indicate the range of the data. (c) Density variations of the Slater determinant states (represented by marker size) along the $[100]$, $[111]$, and $[\bar{1}\bar{1}\bar{1}]$ directions of the Bloch sphere. Dashed lines denote the charge-density-wave supercell.} 
    \label{fig:SU2_CDW}
\end{figure}

\textit{Multicomponent states.—}
We now focus on the $\mathcal{C}=-2$ ideal band in the $N=3$ model and study $\hat{\mathcal{H}}_2$ with the nearest-neighbour interaction. At $\nu = 1/2$, we perform ED calculation in the $N_s=28$ cluster. As shown in Fig.~\ref{fig:SU2_CDW}(a), the ED spectrum reveals a 15-fold quasi-degenerate ground state with a clear energy gap and a quantized many-body Chern number $\mathcal C_{\text{avg}} = -1$. This quasi-degeneracy is consistent with an emergent SU(2)-symmetric ground-state manifold with total spin $S = N_e/2=7$. This SU(2) structure originates from the band properties: the $\mathcal{C}=-2$ band can be folded into two $\mathcal{C}=-1$ subbands with ideal quantum geometry, forming a pseudospin-$1/2$ degree of freedom \cite{Wang2023Origin,Dong2023Manybody}. At $\nu=1/2$, this intrinsic two-component structure gives rise to an integer anomalous Hall crystal state with emergent SU(2) symmetry \cite{Dong2023Manybody,Niu2025Quantum}. %In the continuum limit, the fully filled Slater determinant wave function of an ideal band factorize into a 0LL wavefunction times a periodic envelope, which vanishes identically whenever two particles coalesce. Consequently, the product state is annihilated by the contact interaction $\sum_{i<j}\delta(\bl r_i-\bl r_j)$ and forms an exact zero-energy ground state for short-range repulsive potentials.

To further reveal this ground-state manifold, we construct a Slater determinant state ansatz for the integer anomalous Hall crystal, 
\begin{equation}
\begin{aligned}
\ket{\Xi^{(\alpha,\beta)}}=\frac{1}{S_{\alpha,\beta}}\prod_{\bl k\in\mathrm{hBZ}}\chi^{(\alpha,\beta)\dagger}_{\bl k}\ket{0},
\end{aligned}
\end{equation}
where hBZ denotes the half Brillouin zone spanned by $\bl M_1=\bl G_1/2$ and $\bl G_2$, $\ket{0}$ denotes the vacuum state, and $S_{\alpha,\beta}$ is the normalization factor. Here $\bl G_{1,2}=\frac{4\pi}{3a}\left[\pm \frac{1}{2},\frac{\sqrt{3}}{2}\right]$ are the reciprocal lattice vectors of the original unit cell. $\chi^{(\alpha,\beta)\dagger}_{\bl k}$ is defined as 
\begin{equation}
\label{eq:chi_alpha_beta}
\begin{aligned}
\chi^{(\alpha,\beta)\dagger}_{\bl k} & =h_{\bl k}[\alpha (\varphi_{\bl k}^\dagger+e^{i\ell^2\bl M_1\times \bl k }\varphi_{\bl k+\bl M_1}^\dagger) \\
 & +\beta e^{i\ell^2\bl M_2\times \bl k }(\varphi_{\bl k+\bl M_2}^\dagger-ie^{i\ell^2\bl M_1\times\bl k }\varphi_{\bl k+\bl M_1+\bl M_2}^\dagger)],
\end{aligned}
\end{equation}
where $\varphi_{\bl k}^\dagger=\mathcal N_{0,\bl k}^{-1}\mathcal N_{1,\bl k}^{-1}\Phi_{2,\bl k}^\dagger$, $\Phi_{2,\bl k}^\dagger$ is the creation operator for the Bloch state $\ket{\Phi_{2,\bl k}}$ and $h_{\bl k}$ is the normalization factor. Each term in Eq.~\eqref{eq:chi_alpha_beta} corresponds to a wave function that is an anti-holomorphic function of $\bl k$ multiplied by $h_{\bl k}\exp[-\ell^2\bl k^2/2]$, and therefore, the combination always corresponds to an ideal Chern band. We can parametrize $\alpha= \cos(\theta/2)$ and $\beta= \sin(\theta/2) e^{i \phi}$, where $\theta$ and $\phi$ define, respectively, polar and azimuthal angles of a Bloch sphere.

Figure \ref{fig:SU2_CDW}(b) presents the mean-field (MF) energy $E_{\text{MF}}=\bra{\Xi^{(\alpha,\beta)}}\hat{\mathcal{H}}_2\ket{\Xi^{(\alpha,\beta)}}$ on the Bloch sphere for 14 particles. The smooth energy variation and its proximity to the ED ground-state energies demonstrate the validity of this variational description. The weak energy anisotropy indicates that the emergent SU(2) symmetry is approximate. The states along the $\pm[100]$, $\pm[010]$, and $\pm[001]$ directions are degenerate and correspond to $1\times2$ charge orders related by $C_3$ symmetry, with the charge density along $[100]$ shown in Fig.~\ref{fig:SU2_CDW}(c). The energies are maximized and minimized along the $[111]$ and $[\bar1\bar1\bar1]$ directions, respectively, where the corresponding integer anomalous Hall crystal states exhibit complementary $2\times2$ charge orders with $C_3$ symmetry [Figs.~\ref{fig:SU2_CDW}(d-e)].

\begin{figure}[t]
    \includegraphics[width=1.\columnwidth]{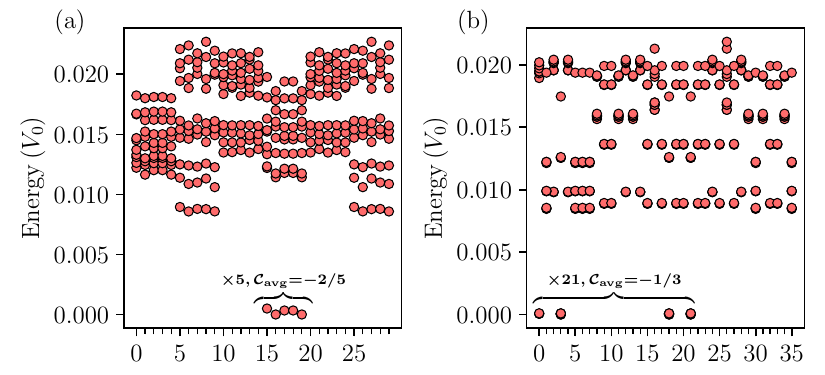}
    \caption{(a,b) ED spectra at $\nu=1/5$ and $1/6$ in the $\mathcal{C}=-2$ band of the $N=3$ model on clusters with $N_s=30$ and $36$, respectively. Energy is measured relative to the ground state.} 
    \label{fig:highC}
\end{figure}

We then perform ED calculations at $\nu=1/5$ and $1/6$ using $N_s=30$ and $36$ clusters, respectively. At $\nu=1/5$, the spectrum [Fig.~\ref{fig:highC}(a)] exhibits a fivefold quasi-degenerate ground-state manifold with $\mathcal C_{\text{avg}} = -2/5$, consistent with the Halperin (332) state generated by vortex attachment to the filled Slater determinant at $\nu=1$ \cite{Dong2023Manybody,Ledwith2023Vortexability,Fujimoto2025Higher}. At $\nu=1/6$, we find a 21-fold quasi-degenerate manifold with $\mathcal C_{\text{avg}} = -1/3$ [Fig.~\ref{fig:highC}(b)], corresponding to a fractional anomalous Hall crystal obtained by attaching vortices to the integer anomalous Hall crystal at $\nu=1/2$. The degeneracy factorizes into a threefold Laughlin-type topological degeneracy and a sevenfold internal multiplicity $(N_e+1)=7$ associated with the emergent $\mathrm{SU}(2)$ structure. The PES shows robust gaps at both fillings, with counting consistent with generalized Pauli principles of type $(1,3)$ and $(1,6)$ for $\nu=1/5$ and $1/6$, respectively \cite{Wang2022Hierarchy,Dong2023Manybody,Liu2025Engineering}. Similarly, in the $\mathcal{C}=-3$ band of the $N=4$ model, ED reveals an integer anomalous Hall crystal, a multicomponent Halperin state, and a fractional anomalous Hall crystal at $\nu=1/3$, $1/7$, and $1/9$, respectively \cite{SM}.

\textit{Discussion.—}
Our lattice models provide a framework for realizing LL physics with a small number of sublattices, and for exploring correlated phases beyond conventional LLs. They are connected to realistic systems: the $N=2$ honeycomb model can be effectively realized in tMoTe$_2$ near the magic angle, while at certain twist angles tMoTe$_2$ admits a three-orbital description involving A, B, and O orbitals with band topology analogous to the $N=3$ model \cite{Qiu2023}. These connections suggest that tMoTe$_2$ \cite{reddy2024nonabelian,Ahn2024NonAbelian,Wang2025Higher,Xu2025Multiple,Chen2025Robust,li2026abelian} and other moir\'e  or skyrmion-lattice \cite{Wang2025Orbital} systems  may serve as platforms for realizing the $N\geq3$ models. Moreover, generalized LL states provide a basis for decomposing Bloch states \cite{liu2025theoryofgLL,li2025Variational,li2026abelian}, offering a guiding principle for identifying lattice systems favorable for fractionalization.

The quantum-geometric properties of our models are robust against hopping truncation: restricting hoppings to distances $d \leq 2a$ preserves the essential band structures and quantum geometry of the original models \cite{SM}. These short-range models could be suitable for quantum simulation platforms that have recently realized fractionalized states, including Floquet-engineered optical lattices with ultracold atoms \cite{Leonard2023Realization} and circuit quantum electrodynamics architectures with photons \cite{Wang2024Realization}. Our results open new opportunities for realizing Abelian, non-Abelian, and multicomponent fractionalized phases in both material and quantum-simulation platforms.

\textit{Acknowledgments.—}We thank Zhao Liu and Jie Wang for valuable discussions, and Junkai Dong for helpful communications.
This work was supported by National Key Research and Development Program of China (Grants No. 2022YFA1402400 and No. 2021YFA1401300), National Natural Science Foundation of China (Grants No. 12274333 and No. 12550404). %The numerical calculations in this paper have been performed on the supercomputing system in the Supercomputing Center of Wuhan University.

\bibliography{ref}

\end{document}

\title{Supplemental Material for ``Quantum-Geometric Design of Lattice Generalized Landau Levels"}

\author{Bohao Li}
\affiliation{School of Physics and Technology, Wuhan University, Wuhan 430072, China}
\author{Fengcheng Wu}
\email{wufcheng@whu.edu.cn}
\affiliation{School of Physics and Technology, Wuhan University, Wuhan 430072, China}
\affiliation{Wuhan Institute of Quantum Technology, Wuhan 430206, China}

\maketitle

\section{magnetic Bloch wave function}

We present a brief review of the magnetic Bloch wave function for the $n$th Landau level (nLL). The magnetic Bloch wave function for 0LL is given by 
 \begin{equation}
 \label{MBS0}
 \begin{aligned}
     \Psi_{0,\bl k}(\bl r)=&\frac{1}{S\ell}\sigma(z+iz_{\bl k}\ell^2 )e^{-\frac{1}{4}\lvert z_{\bl k}\rvert^2\ell^2-\frac{1}{4}\lvert z\rvert^2\ell^{-2}+\frac{i}{2}z_{\bl k}^*z},
 \end{aligned}     
 \end{equation}
where $z = x + iy, z_{\bl{k}} = k_x + ik_y$ , $S$ is a normalization factor, $\ell = \sqrt{\mathcal{A}_0/(2\pi)}$,  $\mathcal{A}_0$ is the area of the (magnetic) unit cell, and $\sigma(z)$ is the modified Weierstrass sigma function \cite{Haldane2018modular} formulated as
\begin{equation}
\begin{aligned}
\label{sigma}
\sigma(z)= ze^{\frac{\eta_1z^2}{z_1}}\frac{\mathcal{\theta}_1(v\mid\tau)}{v\mathcal{\theta}_1'(0\mid\tau)},
\end{aligned}
\end{equation}
where $\mathcal{\theta}_1(v\mid\tau)$ is the Jacobi theta function, $v=z/z_1$, $\eta_1=z_1^*/(4\ell^2)$, $\tau=z_2/z_1$, and $z_j=R_{j,x}+iR_{j,y}$ with $\bl R_{1,2}$ the basis vectors. Expression of $\mathcal{\theta}_1(u\mid\tau)$ is
\begin{align}
\mathcal{\theta}_{1}(u\mid\tau)=-\sum_{n=-\infty}^{+\infty}e^{i\pi\tau(n+\frac{1}{2})^2}e^{2\pi i(n+1/2)(u+1/2)}.
\end{align}
The magnetic Bloch wavefunction for nLL is formulated as
\begin{equation}
 \label{MBSn}
 \begin{aligned}
     \Psi_{n,\bl k}(\bl r)=&\frac{(a^\dagger)^n}{\sqrt{n!}}\Psi_{0,\bl k}(\bl r),
 \end{aligned}     
 \end{equation}
where $a^\dagger$ and $a$ denote, respectively, the raising and lowering operators,
\begin{equation}
\begin{aligned}
&a^\dagger=i\frac{-2\ell\partial_z+z^*\ell^{-1}/2}{\sqrt2},a=i\frac{-2\ell\partial_{z^*}-\ell^{-1}z/2}{\sqrt2}. 
\end{aligned}
\end{equation} 

$\Psi_{n,\bl k}(\bl r)$ satisfies the following magnetic translational symmetry \cite{li2025Variational},
\begin{equation}
\label{eq:trans_sym_r}
\begin{aligned}
\Psi_{n,\bl k}(\bl r+\bl R_i)
=-e^{i\frac{1}{2\ell^2}\bl R_i\times \bl r}e^{i\bl k\cdot \bl R_i}\Psi_{n,\bl k}(\bl r),
\end{aligned}     
\end{equation}
and exhibits the property of position-momentum duailty,
\begin{align}
\label{eq:p-m}
\Psi_{n,\bl k}(\bl r)= &\Psi_{n,-\ell^{-2}\bl e_z\times\bl r}(\ell^{2}\bl e_z\times\bl k) e^{i\bl k\cdot\bl r},
\end{align}
which can be verified from Eqs.~\eqref{MBS0} and \eqref{MBSn}. 
% A corollary of Eqs. \eqref{eq:trans_sym_r} and \eqref{eq:p-m} is that $\Psi_{n,\bl k}(\bl r)$ satisfies
% \begin{equation}
% \begin{small}
% \begin{aligned}
% &\Psi_{n,\bl k+\bl G_i}(\bl r)\\
% =&\Psi_{n,-\ell^{-2}\bl e_z\times\bl r}(\ell^{2}\bl e_z\times\bl k+\bl R_i)e^{i(\bl k+\bl G_i)\cdot\bl r}\\
% =&-
% e^{i\frac{1}{2}\bl R_i\times (e_z\times\bl k)} e^{-i\ell^{-2}\bl R_i\cdot (\bl e_z\times\bl r)}e^{i\bl G_i\cdot\bl r}\Psi_{n,\bl k}(\bl r)\\
% =&-e^{i\frac{\ell^2}{2}\bl G_i \times\bl k}\Psi_{n,\bl k}(\bl r)
% \end{aligned}   
% \end{small}  
% \end{equation}
% where $\bl G_{1,2}=-\ell^{-2}e_z\times\bl R_{1,2}$ are the primitive reciprocal lattice vectors. 
Consequently, $\Psi_{n,\bl k}(\bl r)$ has quasi-periodicity in momentum space \cite{Wang2021Exact},
\begin{equation}
\label{eq:trans_sym_k}
\begin{small}
\begin{aligned}
&\Psi_{n,\bl k+\bl G_i}(\bl r)=-e^{i\frac{\ell^2}{2}\bl G_i \times\bl k}\Psi_{n,\bl k}(\bl r),
\end{aligned}   
\end{small}  
\end{equation}
where $\bl G_{1,2}=\frac{4\pi}{\sqrt{3}a}[\pm\frac{\sqrt{3}}{2},\frac{1}{2}]$ are the primitive reciprocal lattice vectors. 

\section{density-modulated basis}
The construction of the generalized $n$LL band for $0 \le n \le N-2$ in an $N$-sublattice model relies on the assumption that the density-modulated basis states $|e_{n,\bl k}\rangle$ are linearly independent for $0 \le n \le N-2$. For the $N=2,3,$ and $4$ models constructed in the main text, this condition can be verified by explicitly computing the overlap matrix (i.e., Gram matrix) $\langle e_{m,\bl k}|e_{n,\bl k}\rangle$, which remains positive definite throughout the Brillouin zone. 

We note, however, that this condition is not automatically satisfied. A counterexample is the kagome lattice with sublattices A, B, and C located at $\bl\tau_1 = a[1/2, 0]^{\mathsf{T}}$, $\bl\tau_{2,3} = a[\pm1/4, \sqrt{3}/4]^{\mathsf{T}}$. In this case, we find $\Psi_{1,\bl\Gamma}(\bl \tau_i)=0$ for $i=1,2,3$, and consequently $|e_{1,\bl \Gamma}\rangle=0$ for the 1LL wave function, which renders the Gram–Schmidt procedure ill-defined. Thus, the linear independence condition must be verified for each specific realization.

\section{ideal higher Chern band}
We derive the expression of $\ket{\Phi_{N-1,\bl k}}$  and prove that it has ideal quantum geometry. To this end, we first represent $\ket{e_{n,\bl k}}$ and $\ket{\Phi_{n,\bl k}}$ in the orbital Bloch basis $\ket{\psi_{\bl k,s}}$,
\begin{equation}
\begin{aligned}
 \ket{e_{n,\bl k}}=\sum_s\zeta_{n,s}(\bl k)\ket{\psi_{\bl k,s}},\ket{\Phi_{n,\bl k}}=\sum_s\xi_{n,s}(\bl k)\ket{\psi_{\bl k,s}}.
\end{aligned}
\end{equation}
$\ket{\psi_{\bl k,s}}$ is expressed as
\begin{equation}
\begin{aligned}
 \ket{\psi_{\bl k,s}}=\frac{1}{\sqrt{N_d}}\sum_{\bl R}e^{i\bl k\cdot (\bl R+\bl \tau_s)}\ket{\bl R+\bl\tau_s},
\end{aligned}
\end{equation}
where $N_d$ is the number of unit cells and $s$ denotes the sublattice index. $\zeta_{n,s}(\bl k)$ is given by 
\begin{equation}
\begin{aligned}
&\zeta_{n,s}(\bl k)= \braket{\psi_{\bl k,s}}{e_{n,\bl k}}\\
= &\frac{1}{N_d}\sum_{\bl R}\mathcal B(\bl R+\bl \tau_s)\Psi_{n,\bl k}(\bl R+\bl \tau_s)e^{-i\bl k\cdot (\bl R+\bl \tau_s)}\\
= &\mathcal B(\bl\tau_s)u_{n,\bl k}(\bl \tau_s),
\end{aligned}
\end{equation}
where $u_{n,\bl k}(\bl r)=e^{-i\bl k\cdot\bl r}\Psi_{n,\bl k}(\bl r)$. $\xi_{N-1,s}(\bl k)$ is given by
\begin{equation}
\label{eq:N-1_Bloch}
\begin{small}
\begin{aligned}
[\xi_{N-1,s}(\bl k)]^*=\det{\hat{\bl e}_s, \bl\xi_{0}(\bl k), ..., \bl\xi_{N-2}(\bl k)},
\end{aligned}
\end{small}
\end{equation}
where $\hat{\bl e}_s$ is the $s$-th unit vector with $(\hat{\bl e}_s)_m = \delta_{ms}$ and $\bl\xi_{n}(\bl k)=[\xi_{n,1}(\bl k),...,\xi_{n,N}(\bl k)]^\mathsf{T}$. The completeness condition ensures that Eq.~\eqref{eq:N-1_Bloch} determines $\xi_{N-1,s}(\bl k)$ uniquely up to an overall phase factor.

We then prove the ideal quantum geometry of $\ket{\Phi_{N-1,\bl k}}$ by demonstrating that $[\xi_{N-1,s}(\bl k)]^*$ is a holomorpic function  of $z_{\bl k}$ up to a normalization factor. $\ket{\Phi_{n,\bl k}}$ for $0\le n\le N-2$ are obtained from $\ket{e_{n,\bl k}}$ via Gram-Schmidt orthogonalization. Therefore, we obtain the linear transformation from ${\bl\zeta}_{n}$ to $\bl \xi_m$ for $0\le m\le N-2$,
\begin{equation}
\label{eq:trans1}
\begin{aligned}
\xi_{m,s}(\bl k)=&\sum_{n=0}^{N-2}[U(\bl k)]_{m,n}\zeta_{n,s}(\bl k),
\end{aligned}
\end{equation}
where $U(\bl k)$ is a lower-triangular matrix with positive real diagonal entries. Consequently, we have
\begin{equation}
\begin{aligned}
[\xi_{N-1,s}(\bl k)]^*=&\det [U(\bl k)]\det{\hat{\bl e}_s, {\bl\zeta}_{0}(\bl k), ..., {\bl\zeta}_{N-2}(\bl k)},
\end{aligned}
\end{equation}
where $\det [U(\bl k)]$ is a real and positive function of $\bl k$. 

We note that $u_{n,\bl{k}}(\bl{r})$ can be expanded in terms of the $n$th order real-space derivatives of $u_{0,\bl{k}}(\bl{r})$ as follows,
\begin{equation}
\small
\begin{aligned}  
&u_{n,\bl k}(\bl r)=e^{-i\bl k\cdot\bl r}\frac{(a^\dagger)^n}{\sqrt{n!}}\Psi_{0,\bl k}(\bl r)\\
% =\frac{(a^\dagger+\frac{\ell}{\sqrt{2}}z_{\bl k}^*)^n}{\sqrt{n!}}u_{0,\bl k}(\bl r)\\
&=\frac{\ell^n}{\sqrt{2^nn!}}(-2i\partial_z+\frac{iz^*}{2\ell^2}+z_{\bl k}^*)^nu_{0,\bl k}(\bl r)\\
&=\gamma_{n}\ell^n\partial_{z}^{n}u_{0,\bl k}(\bl r)+\sum_{m=0}^{n-1}\beta_{m,n}(\bl k,\bl r)\ell^m\partial_{z}^{m}u_{0,\bl k}(\bl r),
\end{aligned}
\end{equation}
where $\beta_{m,n}(\bl k,\bl r)$ for $0\le m\le n-1$ are functions of $\bl k$ and $\bl r$,  and $\gamma_{n}=\frac{(-\sqrt{2}i)^n}{\sqrt{n!}}$ is independent of $\bl k$. We then obtain the following expression,
\begin{equation}
\label{eq:trans1}
\begin{aligned}
&\zeta_{m,s}(\bl k)=\mathcal B(\bl\tau_s)u_{m,\bl k}(\bl \tau_s)=\sum_{n=0}^{N-2}[\wt U(\bl k)]_{m,n}\wt\zeta_{n,s}(\bl k)\\
&\wt\zeta_{n,s}(\bl k)=B(\bl\tau_s)\ell^n\partial_{z}^{n}u_{0,\bl k}(\bl \tau_s),
\end{aligned}
\end{equation}
where $\wt U(\bl k)$ is a lower-triangular matrix and $[\wt U(\bl k)]_{n,n}=\gamma_n$. Therefore, $C_0=\det [\wt U(\bl k)] = \Pi_{n=0}^{N-2} \gamma_n$ is independent of $\bl k$. $\wt{\bl\zeta}_{n}(\bl k)$ is equivalent to a holomorphic function $\bl\mu_n(\bl k)$ in momentum space up to a normalization factor,
\begin{equation}
\small
\begin{aligned}
&\wt{\zeta}_{n,s}(\bl k)=e^{\frac{\ell^2}{4}|z_{\bl k}|^2}\mu_{n,s}(\bl k),\\
&\mu_{n,s}(\bl k)=e^{-\frac{\ell^2}{4}|z_{\bl k}|^2}B(\bl\tau_s)\ell^n\partial_{z}^{n}u_{0,\bl k}(\bl \tau_s).
\end{aligned}
\end{equation}
$[\xi_{N-1,s}(\bl k)]^*$ thus has momentum-space holomorphicity up to a normalization factor,
\begin{equation}
\begin{aligned}
[\xi_{N-1,s}(\bl k)]^*=&C_0\det [U(\bl k)]\det{\hat{\bl e}_s, \wt{\bl\zeta}_{0}(\bl k), ..., \wt{\bl\zeta}_{N-2}(\bl k)}\\
=& \det [U(\bl k)]e^{-\frac{N-1}{4}\ell^2|z_{\bl k}|^2}g_{s}(z_{\bl k}),
\end{aligned}
\label{eq:decomposition_highC}
\end{equation}
where $g_{s}(z_{\bl k})$ is a holomorpic function of $\bl z_{\bl k}$,
\begin{equation}
\begin{aligned}
g_{s}(z_{\bl k}) =& C_0\det{\hat{\bl e}_s, \bl\mu_{0}(\bl k), ..., \bl\mu_{N-2}(\bl k)}.
\end{aligned}
\end{equation}
The ideal trace condition of $\ket{\Phi_{N-1,\bl k}}$ is then satisfied \cite{Claassen2015,Wang2021Exact}.

\section{hopping parameters}
The hopping parameters, obtained via the Fourier transform of the the Hamiltonian matrix elements $(H_{\bl k})_{i,j}=\bra{\psi_{\bl k,i}}\hat H\ket{\psi_{\bl k,j}}$, are given by
\begin{equation}
\label{eq:Jij}
\begin{small}
\begin{aligned}
J_{ij}(\bl R)= & \mathcal A_0\int_{\mathcal M_{\mathrm{BZ}}}\frac{d^2\bl k}{(2\pi)^2}(H_{\bl k})_{i,j}e^{i\bl k\cdot (\bl R+\bl \tau_i-\bl \tau_j)},
\end{aligned}
\end{small}
\end{equation}
where $\mathcal M_{\mathrm{BZ}}$ denotes the Brillouin zone.
 
Here we analytically derive the hopping parameters for the Gaussian decay models with a zero-energy band as the lattice counterpart of the generalized 0LL, to which the $N = 2$ model in the main text belongs. We also demonstrate that the Gaussian decay model is equivalent to the generalized Kapit-Mueller model in non-Bravais lattices up to a site-dependent gauge transformation \cite{Dong2020Exact}. We finally discuss the $N =3$ and $4$ models constructed in the main text.

\subsection{Gaussian-decay model} 
We define the Gaussian-decay models with $N$ sublattices by taking $\mathcal E_{0,\bl k}=0$ and $\mathcal E_{n,\bl k}=\mathcal{N}_{\bl k,0}^{-2}\mathcal A_0$ for $1\le n\le N-1$, which yields analytic, Gaussian-decaying hopping amplitudes in real space. Here we set the energy to be dimensionless. We note that this Gaussian-decay model on the honeycomb lattice is the $N=2$ model in the main text. The matrix elements for the Gaussian-decay models are given by
\begin{equation}
\begin{small}
\begin{aligned}
&\mathcal A_0^{-1}(H_{\bl k})_{i,j}\\
=&-\mathcal B(\bl \tau_i)u_{0,\bl k}(\bl \tau_i)[\mathcal B(\bl \tau_j)u_{0,\bl k}(\bl \tau_j)]^*\!+\!\delta_{i,j}\!\sum_{s}\!|\mathcal B(\bl \tau_s)|^2|u_{0,\bl k}(\bl \tau_s)|^2.
\end{aligned}
\end{small}
\end{equation}
$J_{ij}(\bl R)$ is then obtained by
\begin{equation}
\label{eq:J_Gaussian}
\begin{small}
\begin{aligned}
&\frac{(2\pi)^2}{\mathcal A_0^2}J_{ij}(\bl R)\\
=&-\mathcal B(\bl \tau_i)[\mathcal B(\bl \tau_j)]^*f(\bl \tau_j,\bl \tau_i,\bl R)+\delta_{i,j}\!\sum_s\!|\mathcal B(\bl \tau_s)|^2f(\bl \tau_s,\bl \tau_s,\bl R),
\end{aligned}
\end{small}
\end{equation}
where $f(\bl \tau,\bl \tau',\bl R)$ represents the LL matrix element in momentum space,
\begin{equation}
\begin{small}
\begin{aligned}
f(\bl \tau,\bl \tau',\bl R)=\int_{\mathcal M_{\mathrm{BZ}}} d^2\bl k[\Psi_{0,\bl k}(\bl \tau)]^*\Psi_{0,\bl k}(\bl \tau')e^{i\bl k\cdot \bl R}.
\end{aligned}
\end{small}
\end{equation}
This function can be rewritten using Eq.~\eqref{eq:p-m},
\begin{equation}
\begin{small}
\begin{aligned}
&f(\bl \tau,\bl \tau',\bl R)\\
=&\int_{\mathcal M_{\mathrm{BZ}}} d^2\bl k[\Psi_{0,-\ell^{-2}\bl e_z\times\bl \tau}(\ell^{2}\bl e_z\times\bl k)]^*\Psi_{0,-\ell^{-2}\bl e_z\times\bl \tau'}(\ell^{2}\bl e_z\times\bl k)\\
&\times e^{i\bl k\cdot (-\bl \tau+\bl \tau'+\bl R)} \\
= & \ell^{-4}\int_{\mathcal M_{\mathrm{UC}}} d^2\bl r[\Psi_{0,\bl q}(\bl r)]^*\Psi_{0,\bl q'}(\bl r)e^{i(\bl q-\bl q'+\bl Q)\cdot\bl r},
\end{aligned}
\end{small}
\end{equation}
where $\bl q=-\ell^{-2}\bl e_z\times\bl \tau$, $\bl q'=-\ell^{-2}\bl e_z\times\bl \tau'$, $\bl Q=\ell^{-2}\bl e_z\times\bl R$ is a reciprocal lattice vector, and $\mathcal M_{\mathrm{UC}}=\{\bl r|\bl r=\ell^{2}\bl e_z\times\bl k,\bl k\in\mathcal M_{\mathrm{BZ}}\}$ is the Wigner-Seitz unit cell in real space. Given the definition of the LL matrix element in real space,
\begin{equation}
\begin{aligned}
\small
\wt f(\bl q,\bl q',\bl k)=\int_{\mathcal M_{\mathrm{UC}}} d^2\bl r[\Psi_{0,\bl q}(\bl r)]^*\Psi_{0,\bl q'}(\bl r)e^{i\bl k\cdot\bl r},
\end{aligned}
\end{equation}
we have
\begin{equation}
\begin{aligned}
\small
f(\bl \tau,\bl \tau',\bl R)=\ell^{-4}\wt f(\bl q,\bl q',\bl q-\bl q'+\bl Q).
\end{aligned}
\end{equation}
Using $\int_{\mathcal M_{\mathrm{UC}}} d^2\bl r\;\lvert\Psi_{0,\bl k}(\bl r)\rvert^2=1$, the expression of $\wt f(\bl q,\bl q',\bl q-\bl q'+\bl Q)$ is given by \cite{Wang2021Exact}
\begin{equation}
\begin{aligned}
\small
&\wt f(\bl q,\bl q',\bl q-\bl q'+\bl Q)\\
&= e^{i\phi(\bl Q)} e^{-\frac{i\ell^2}{2}(\bl q+\bl q') \times \bl Q} e^{\frac{i\ell^2}{2}\bl q \times \bl q'} e^{-\frac{1}{4}|\bl q-\bl q'+\bl Q|^2\ell^2}.
\end{aligned}
\end{equation}
where $\phi(\bl Q)=m+n+mn$ for $\bl Q=m\bl Q_1+n\bl Q_2$, and $\bl Q_{1},\bl Q_{2}$ are the reciprocal lattice basis vectors. Therefore $f(\bl \tau,\bl \tau',\bl R)$ is given by
\begin{equation}
\label{eq:J_N=2}
\begin{aligned}
\small
&f(\bl \tau,\bl \tau',\bl R)\\
&= \ell^{-4}\eta_{\bl R} e^{\frac{i}{2\ell^{2}}(\bl \tau+\bl \tau') \times \bl R} e^{\frac{i}{2\ell^{2}}\bl \tau \times \bl \tau'} e^{-\frac{1}{4\ell^{2}}|-\bl \tau+\bl \tau'+\bl R|^2}.
\end{aligned}
\end{equation}
where $\eta_{\bl R}=(-1)^{m+n+mn}$ for $\bl R=m\bl R_1+n\bl R_2$. 

\begin{figure*}[p]   
\includegraphics[width=1.\columnwidth]{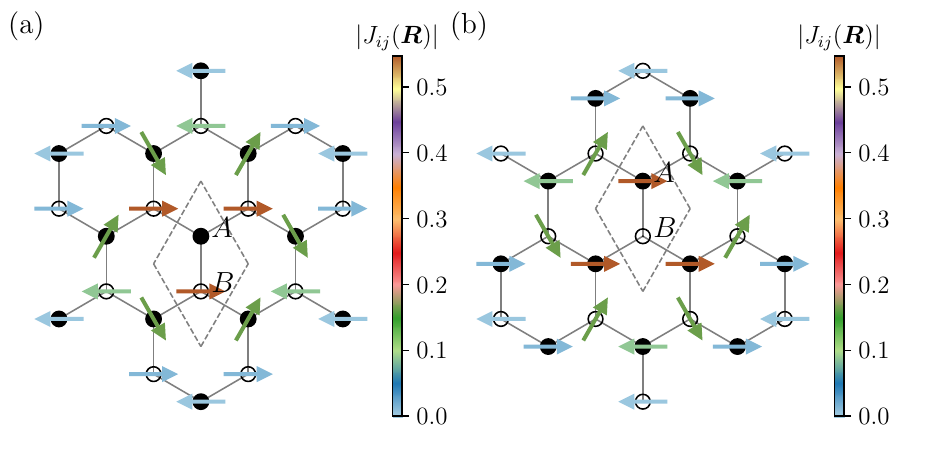}
\caption{Hopping parameters (a) $J_{iA}(\bl R)$ and (b) $J_{iB}(\bl R)$ for $N=2$ model in honeycomb lattice. For each arrow, its direction corresponds to the phase factor and the color to the magnitude. Dashed lines outline the unit cell.} 
\label{fig:J_lat_AB}
\end{figure*}

Combining Eqs.~\eqref{eq:J_Gaussian} and \eqref{eq:J_N=2}, we obtain an analytic expression for the hopping parameters that decay as a Gaussian with distance. Applied to the honeycomb lattice, this yields the hopping parameters given in Eq.~(7) of the main text, which are illustrated in Fig.~\ref{fig:J_lat_AB}. From the patterns of the hopping parameters, it is clear that both the A and B sites respect $C_{3}$ symmetry, where $C_n$ represents $n$-fold rotational symmetry around $z$-axis. The hopping terms within the third nearest neighbors are dominant. Truncating to the next-nearest-neighbour hoppings recovers the Haldane model \cite{Haldane1988Model}.

\begin{figure*}[p]   
\includegraphics[width=1.6\columnwidth]{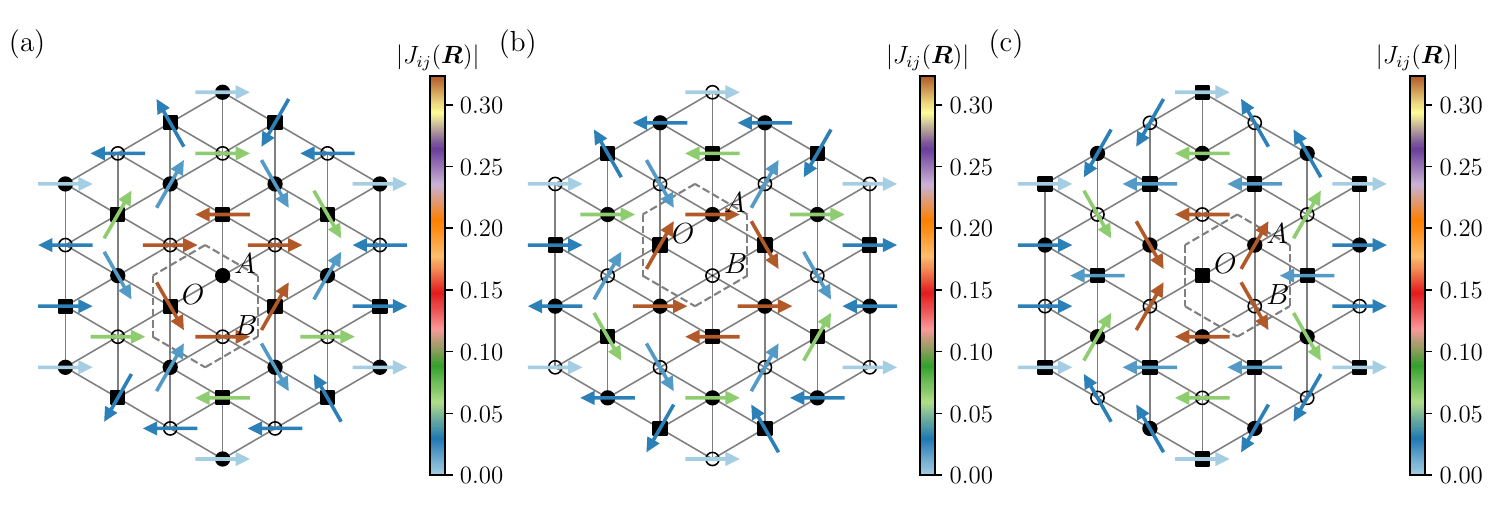}
\caption{Hopping parameters (a) $J_{iA}(\bl R)$, (b) $J_{iB}(\bl R)$, and (c) $J_{iO}(\bl R)$ for $N=3$ model. For each arrow, the direction corresponds to the phase factor and the color to the magnitude. Dashed lines outline the unit cell.} 
\label{fig:J_lat_ABO}
\end{figure*}

\begin{figure*}[p]   
\includegraphics[width=1.4\columnwidth]{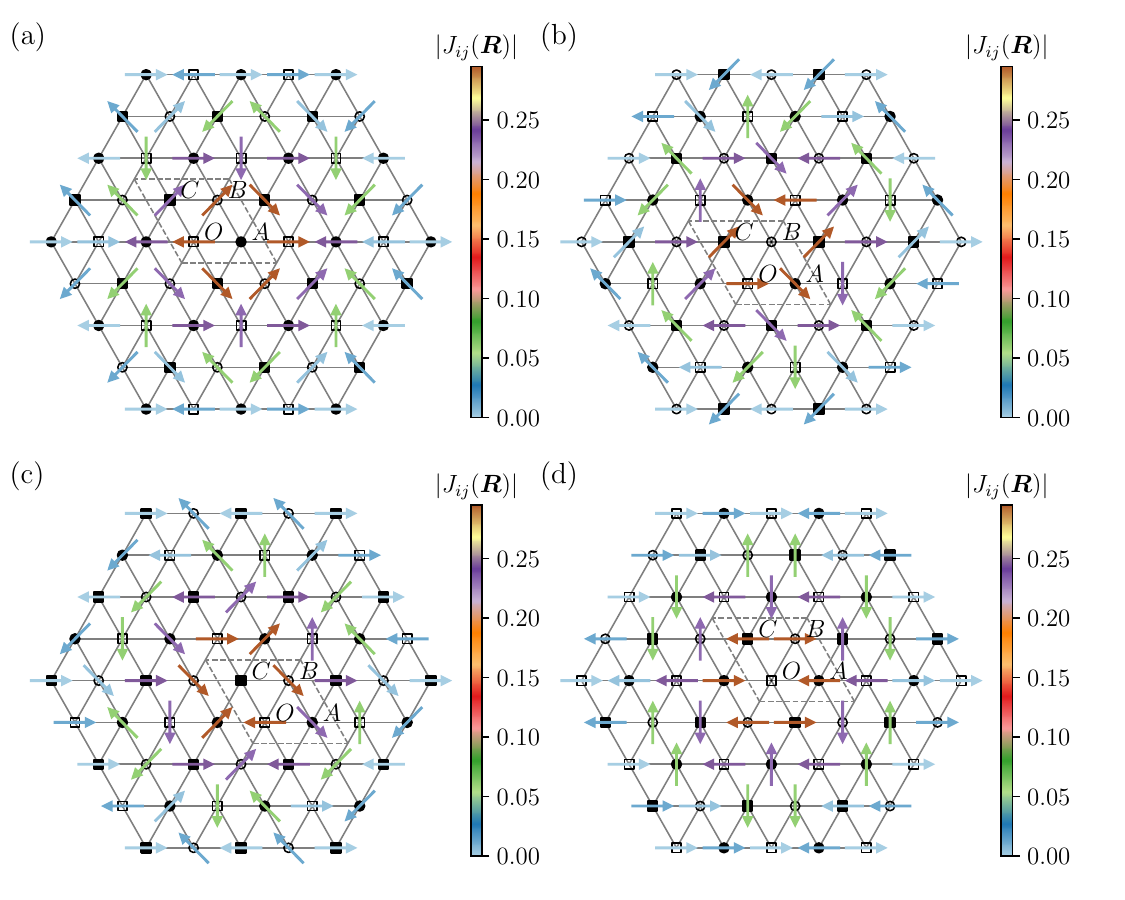}
\caption{Hopping parameters (a) $J_{iA}(\bl R)$, (b) $J_{iB}(\bl R)$, (c) $J_{iC}(\bl R)$, and (d) $J_{iO}(\bl R)$ for $N=4$ model. For each arrow, the direction corresponds to the phase factor and the color to the magnitude. Dashed lines outline the unit cell.} 
\label{fig:J_lat_ABCO}
\end{figure*}
% \clearpage

\begin{figure*}[t]   
\includegraphics[width=2.0\columnwidth]{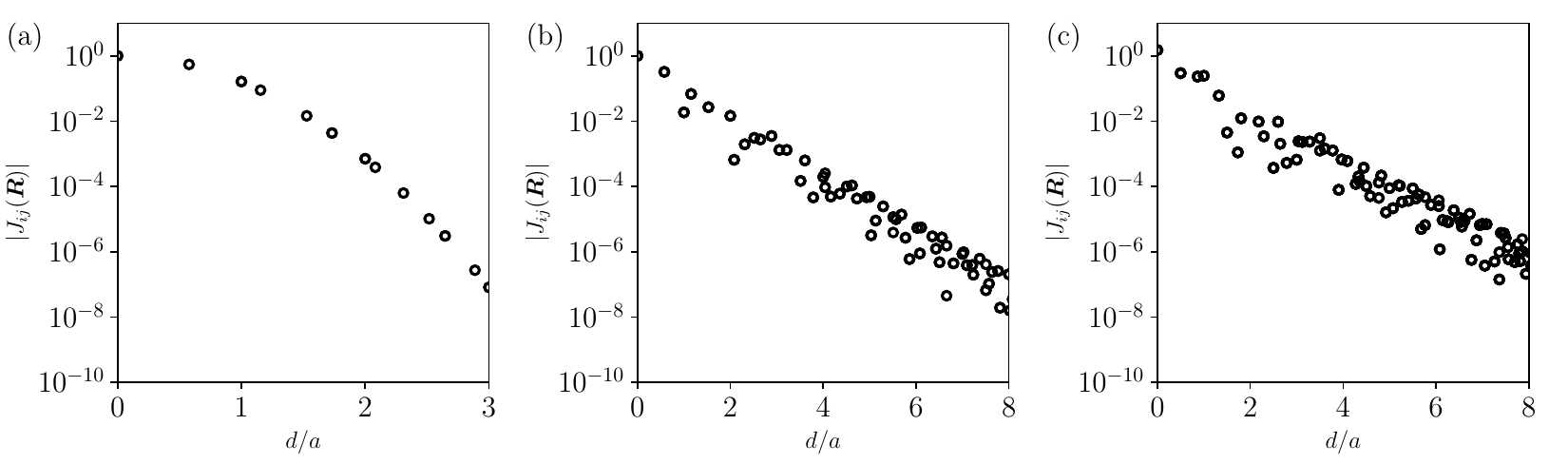}
\caption{$|J_{ij}(\bl R)|$ in a logarithmic scale as a function of $d = |\bl R + \bl\tau_i - \bl\tau_j|$ for the (a) $N=2$, (b) $N=3$, and (c) $N=4$ models.} 
\label{fig:hopping_log}
\end{figure*}

\begin{figure*}[t]   
\includegraphics[width=1.75\columnwidth]{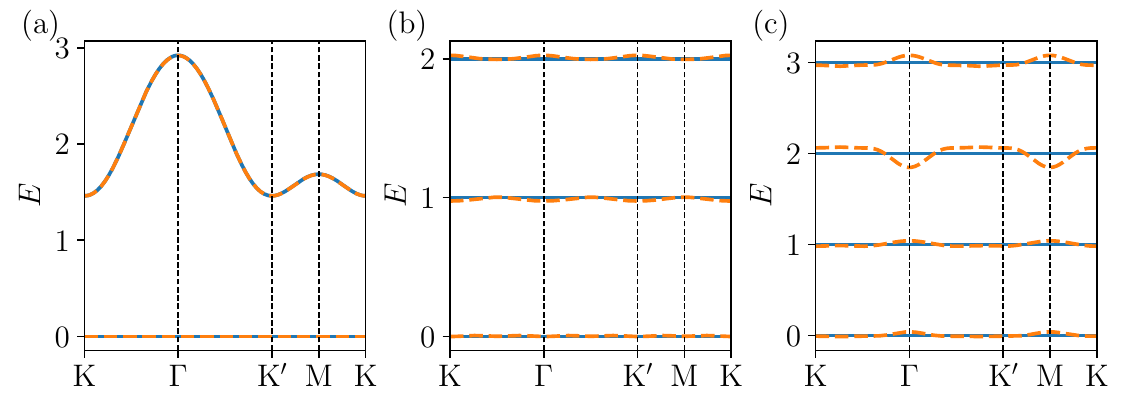}
\caption{Band structures of the original models (blue lines) and the truncated models (orange dashed lines) for $N=2$, $3$, and $4$. In the truncated model, the hopping amplitudes are restricted to distances not exceeding the cutoff $d = 2a$.} 
\label{fig:truncated_bands}
\end{figure*}

\subsection{Generalized Kapit-Mueller model} 
We now present the connection between the Gaussian-decay model and the generalized Kapit-Mueller model in non-Bravais lattices. For the latter model, we focus on the case of one flux quantum per unit cell, of which the Hamiltonian is given by
\begin{equation}
\begin{aligned}
\small
&\hat H^{\mathrm{KM}}=\sum_{\bl R,\bl R'}\sum_{i,j} J_{ij}^{\mathrm{KM}}(\bl R-\bl R') \,
\ket{\bl R+\bl\tau_i}\bra{\bl R'+\bl\tau_j}.
\end{aligned}
\end{equation}
The hopping parameters are
\begin{equation}
\begin{aligned}
\small
&J_{ij}^{\mathrm{KM}}(\bl{R},\bl{R}')\\
&= f_{ij}(\bl{r}_{ij}) \, \exp\!\left[-i\frac{1}{2\ell^2}(\bl{R}+\bl{\tau}_i) \times \bl{r}_{ij}\right] \, \exp\!\left(\frac{1}{4\ell^2}|\bl{r}_{ij}|^2\right),
\end{aligned}
\end{equation}
where $\bl{r}_{ij} = \bl{R}' + \bl{\tau}_j - \bl{R} - \bl{\tau}_i$ and $i$, $j$ denote the sublattice indexes. The Kapit-Mueller model requires that the Hamiltonian annihilates the 0LL wavefunctions as zero modes, which imposes constraints for an arbitrary complex number $c$ and $j=0,1,...,n-1$,
\begin{equation}
\sum_i\sum_{\bl{R}} f_{ji}(\bl{R}+\bl{\tau}_i-\bl{\tau}_j) \, e^{c(z_{\bl{R}} + z_{\bl{\tau}_i} - z_{\bl{\tau}_j})} = 0.
\end{equation}
A nontrivial solution to these constraints is given by \cite{Dong2020Exact}
\begin{equation}
f_{ij}(\bl{r}) = e^{-\frac{1}{2\ell^2}|\bl{r}|^2}\left[1-\delta_{i,j}\sum_{k} e^{i\frac{1}{\ell^2}\,\bl{r} \times (\bl{\tau}_k - \bl{\tau}_i)}\right].
\end{equation}

The Kapit–Mueller model and the Gaussian-decay model are connected via the gauge transformation $|\bl R+\bl\tau_i\rangle \to \mathcal B(\bl R+\bl\tau_i) |\bl R+\bl\tau_i\rangle$, where $|B(\bl R+\bl\tau_i)|=1$. The gauge-transformed Kapit–Mueller Hamiltonian is 
\begin{equation}
\begin{aligned}
&\hat{\mathcal H}_0' = \sum_{\bl R,\bl R'}\sum_{i,j} J_{ij}'(\bl R,\bl R') \ket{\bl R+\bl \tau_i}\bra{\bl R'+\bl \tau_j}, \\
&J_{ij}'(\bl R,\bl R') = J_{ij}^{\mathrm{KM}}(\bl R,\bl R') \, \mathcal{B}(\bl R+\bl \tau_i) \, \mathcal{B}^*(\bl R'+\bl \tau_j),
\end{aligned}
\end{equation}
$J_{ij}'(\bl R,\bl R')$ can be rewritten as
\begin{equation}
\begin{aligned}
\small
& J_{ij}'(\bl R,\bl R') 
% &= J_{ij}^{\mathrm{KM}}(\bl R,\bl R')  \\
% &\times \mathcal{B}(\bl \tau_i) \mathcal{B}^*(\bl \tau_j)\eta_{\bl R-\bl R'}  e^{-\frac{i}{2\ell^2}\bl R\times\bl R'}  e^{-i\frac{1}{2\ell^2}\bl R\times\bl \tau_i}  e^{i\frac{1}{2\ell^2}\bl R'\times\bl \tau_j}\\
% &= \; f_{ij}(\bl{r}_{ij})  e^{\frac{1}{4\ell^2}|\bl{r}_{ij}|^2} 
% e^{-i\frac{1}{2\ell^2}(\bl{R}+\bl{\tau}_i)\times(\bl{R}'+\bl{\tau}_j)} \\
% & \times \mathcal{B}(\bl{\tau}_i) \mathcal{B}^*(\bl{\tau}_j) 
% \eta_{\bl R-\bl R'} 
% e^{-\frac{i}{2\ell^2}\bl{R}\times\bl{R}'} 
% e^{-i\frac{1}{2\ell^2}\bl{R}\times\bl{\tau}_i} 
% e^{i\frac{1}{2\ell^2}\bl{R}'\times\bl{\tau}_j} \\
= \; f_{ij}(\bl{r}_{ij})  e^{\frac{1}{4\ell^2}|\bl{r}_{ij}|^2} 
\eta_{\bl R-\bl R'} \\
&\times\mathcal{B}(\bl{\tau}_i) \mathcal{B}^*(\bl{\tau}_j) 
e^{-i\frac{1}{2\ell^2}(\bl{R}-\bl{R}') \times (\bl{\tau}_i + \bl{\tau}_j)} 
e^{-i\frac{1}{2\ell^2}\bl{\tau}_i \times \bl{\tau}_j},
\end{aligned}
\end{equation}
where we use the identities
\begin{equation}
\begin{aligned}
\mathcal{B}(\bl R+\bl \tau_i) &= \eta_{\bl R} \, e^{-i\frac{1}{2\ell^2}\bl R\times\bl \tau_i} \, \mathcal{B}(\bl \tau_i),
\end{aligned}
\end{equation}
and 
\begin{equation}
\eta_{\bl R}\,\eta_{\bl R'} = \eta_{\bl R-\bl R'} \, e^{-\frac{i}{2\ell^2}\bl R\times\bl R'}.
\end{equation}
We have
\begin{equation}
\label{eq:J_gauge}
\begin{aligned}
\small
&J_{ij}'(\bl R,\bl R') \\
= &\left[1-\delta_{i,j}\sum_{k} e^{i\frac{1}{\ell^2}(\bl R'-\bl R)\times (\bl{\tau}_k - \bl{\tau}_i)}\right] e^{-\frac{1}{4\ell^2}|\bl{r}_{ij}|^2} 
\eta_{\bl R-\bl R'} \\
&\times\mathcal{B}(\bl{\tau}_i) \mathcal{B}^*(\bl{\tau}_j) 
e^{-i\frac{1}{2\ell^2}(\bl{R}-\bl{R}') \times (\bl{\tau}_i + \bl{\tau}_j)} 
e^{-i\frac{1}{2\ell^2}\bl{\tau}_i \times \bl{\tau}_j}, \\
= &\mathcal{B}(\bl{\tau}_i) \mathcal{B}^*(\bl{\tau}_j) e^{-\frac{1}{4\ell^2}|\bl{r}_{ij}|^2} e^{-i\frac{1}{2\ell^2}\overline{\bl R}\times (\bl{\tau}_i + \bl{\tau}_j)}
e^{-i\frac{1}{2\ell^2}\bl{\tau}_i \times \bl{\tau}_j}\eta_{\overline{\bl R}} \\
&- \delta_{i,j} |\mathcal{B}(\bl{\tau}_i)|^2 \sum_{k} e^{-i\frac{1}{\ell^2}\overline{\bl R}\times \bl{\tau}_k } e^{-\frac{1}{4\ell^2}|\overline{\bl R}|^2} 
\eta_{\overline{\bl R}},
\end{aligned}
\end{equation}
where $\overline{\bl R}=\bl{R}-\bl{R}'$. Equation~\eqref{eq:J_gauge} is equivalent to Eq.~\eqref{eq:J_Gaussian} up to a sign.

% For $N=2$ model, we choose $\mathcal B(\bl \tau_1)=1,\mathcal B(\bl \tau_2)=-e^{-\frac{i\ell^{-2}}{2}\bl \tau_1\times\bl \tau_2}$ so that the nearest-neighbour hopping is real and positive, where $J_{ij}(\bl R)$ are formulated in Eq. 7 in the main text. 
% \begin{equation}
%  \label{J_final}
% \begin{small}
% \begin{aligned}
% J_{11}(\bl R)= & \eta_{\bl R} e^{i\ell^{-2}\bl \tau_2 \times \bl R}  e^{-\frac{1}{4}|\bl R|^2\ell^{-2}}\\
% J_{22}(\bl R)= & \eta_{\bl R} e^{i\ell^{-2}\bl \tau_1 \times \bl R}  e^{-\frac{1}{4}|\bl R|^2\ell^{-2}}\\
% J_{12}(\bl R)= & \eta_{\bl R} e^{\frac{i\ell^{-2}}{2}(\bl \tau_1+\bl \tau_2) \times \bl R} e^{-\frac{1}{4}|-\bl \tau_2+\bl \tau_1+\bl R|^2\ell^{-2}}.
% \end{aligned}
% \end{small}
% \end{equation}

\begin{figure}[t]   
\includegraphics[width=1.\columnwidth]{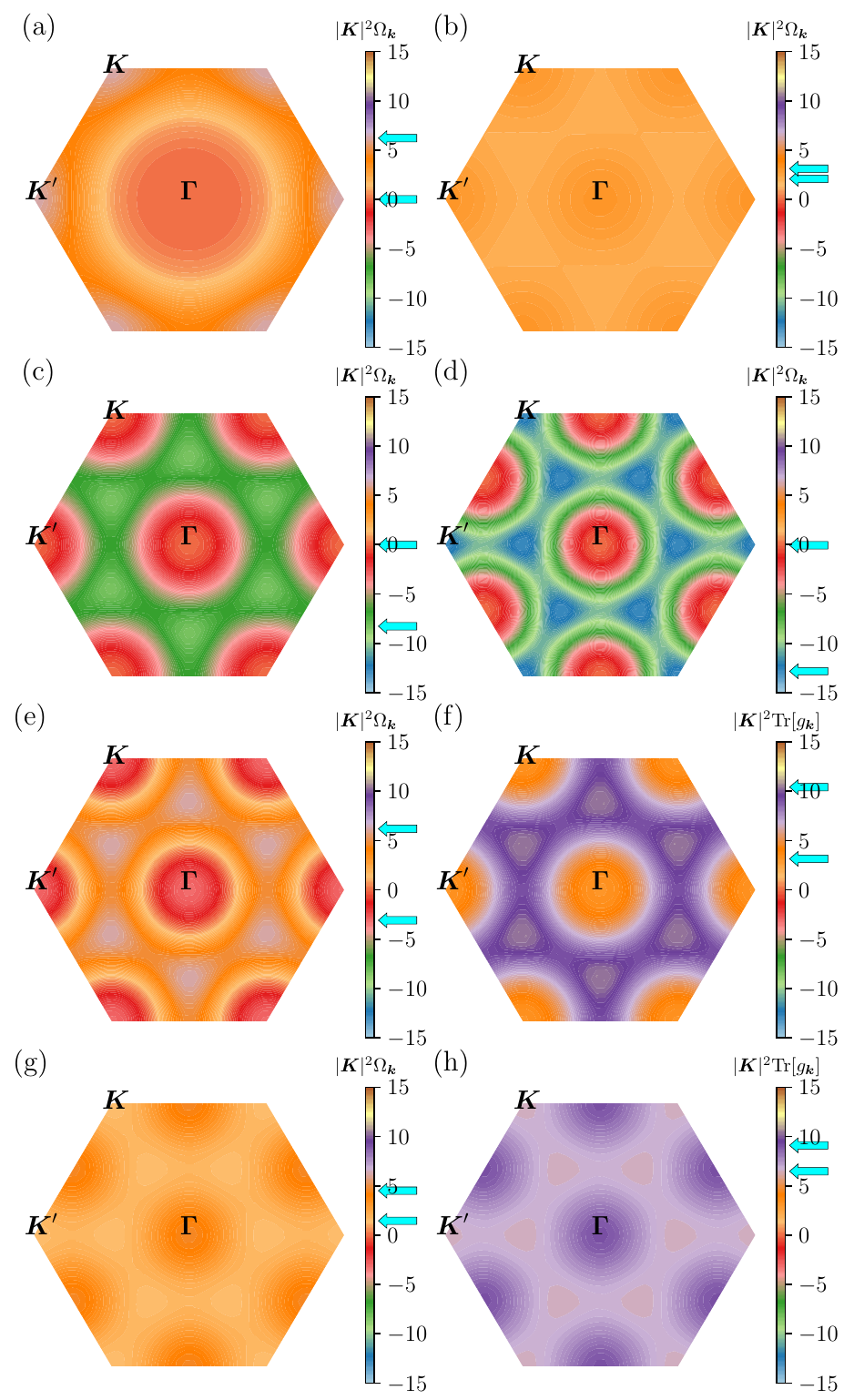}
\caption{(a-b) $\Omega_{\bl k}$($\mathrm{Tr}[g_{\bl k}]$) for the generalized 0LL in the $N=2$ and $N=3$ models. (c-d) $\Omega_{\bl k}$($-\mathrm{Tr}[g_{\bl k}]$) for the higher Chern bands in the $N=3$ and $N=4$ models. (e-f) $\Omega_{\bl k}$ and $\mathrm{Tr}[g_{\bl k}]$ for the generalized 1LL in the $N=3$ model. (g-h) $\Omega_{\bl k}$ and $\mathrm{Tr}[g_{\bl k}]$ for the generalized 1LL in the $N=4$ model. The arrows indicate the data range in each plot.} 
\label{fig:cur_trg}
\end{figure}

\begin{figure}[t]   
\includegraphics[width=1.\columnwidth]{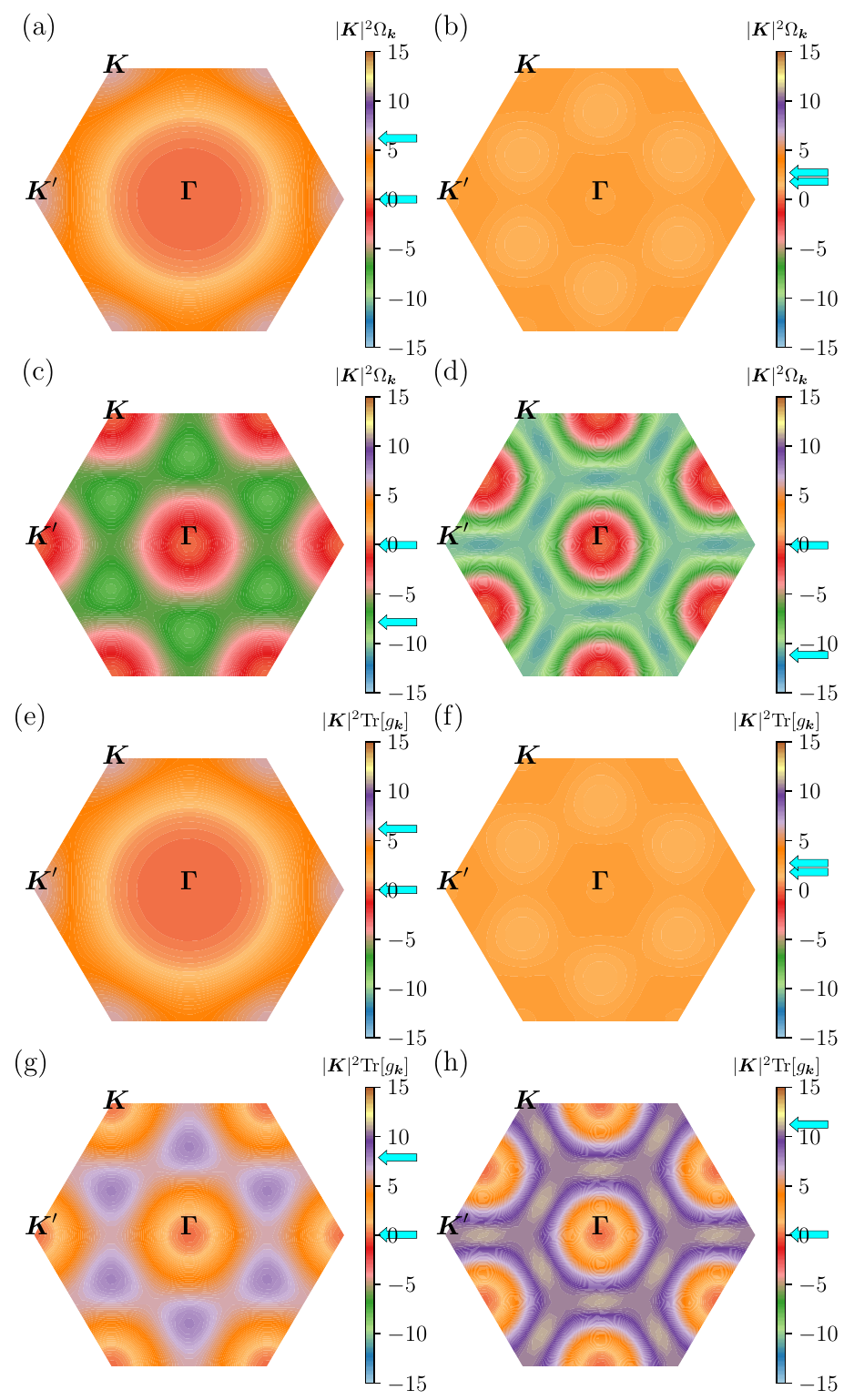}
\caption{(a-b) $\Omega_{\bl k}$ for the generalized 0LL in the truncated $N=2$ and $N=3$ models. (c-d) $\Omega_{\bl k}$ for the higher Chern bands in the truncated $N=3$ and $N=4$ models. (e-f) $\mathrm{Tr}[g_{\bl k}]$ for the generalized 0LL in the truncated $N=2$ and $N=3$ models. (g-h) $\mathrm{Tr}[g_{\bl k}]$ for the higher Chern bands in the truncated $N=3$ and $N=4$ models. The arrows indicate the data range in each plot.} 
\label{fig:cur_trg_trunc1}
\end{figure}

\begin{figure}[t]   
\includegraphics[width=1.\columnwidth]{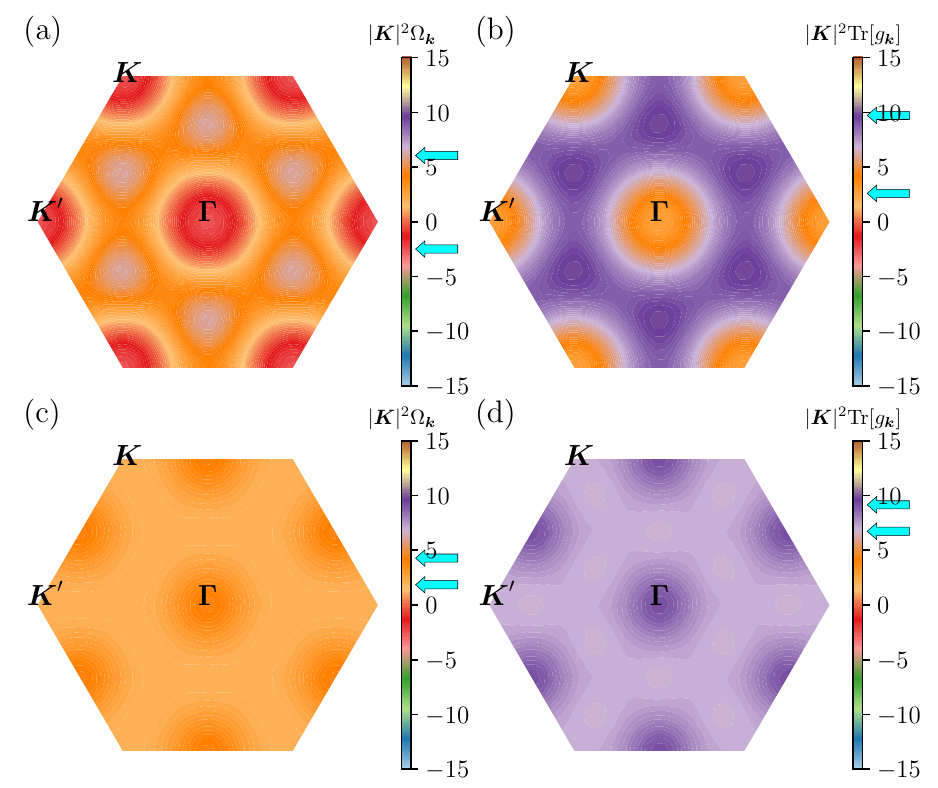}
\caption{(a-b) $\Omega_{\bl k}$ and $\mathrm{Tr}[g_{\bl k}]$ for the generalized 1LL in the truncated $N=3$ model. (c-d) $\Omega_{\bl k}$ and $\mathrm{Tr}[g_{\bl k}]$ for the generalized 1LL in the truncated $N=4$ model. The arrows indicate the data range in each plot.} 
\label{fig:cur_trg_trunc2}
\end{figure}

\subsection{$N$= 3 and 4 models}
The Gaussian-decay model hosts the generalized 0LL as its ground state, while all higher states remain degenerate. To obtain isolated generalized $n$LL bands, we therefore construct the $N=3$ and $4$ models introduced in the main text. The hopping parameters in the $N=3$ and $4$ models, obtained from Eq.~\eqref{eq:Jij}, are shown in Figs. \ref{fig:J_lat_ABO} and \ref{fig:J_lat_ABCO}, respectively. 

In the $N=3$ model, we take $\mathcal B(\bl\tau_1)=\exp(-i\frac{\pi}{3})$, $\mathcal B(\bl\tau_2)=-1$ and $\mathcal B(\bl\tau_3)=\exp(i\frac{\pi}{3})$ such that the A ($\bl\tau_1$) and B ($\bl\tau_2$) sites possess an explicit threefold rotational symmetry, while the O ($\bl\tau_3$) site at the honeycomb center exhibits $C_{6}$ symmetry. The nearest-neighbor and third-nearest-neighbor hoppings are the most dominant, with the next-nearest-neighbor hopping being substantially smaller. 

In the $N=4$ model, we choose $\mathcal B(\bl\tau_i)=1$ for $i=1$ (A), $3$ (C), $4$ (O) and $\mathcal B(\bl\tau_2)=-1$ (B). The A, B, and C sites each host $C_2$ symmetry, whereas the O site retains $C_{6}$ point group symmetry. In this model, the first four hopping terms are dominant.

Figure~\ref{fig:hopping_log} shows $|J_{ij}(\bl R)|$ as a function of hopping distance on a logarithmic scale for the $N=2$, $3$, and $4$ models. In the $N=3$ and $4$ models, $|J_{ij}(\bl R)|$ exhibits exponential decay, while $|J_{ij}(\bl R)|$ exhibits a Gaussian-decay for the $N=2$ model.

\section{quantum geometry and truncated models}

\begin{figure}[t]   
\includegraphics[width=1.\columnwidth]{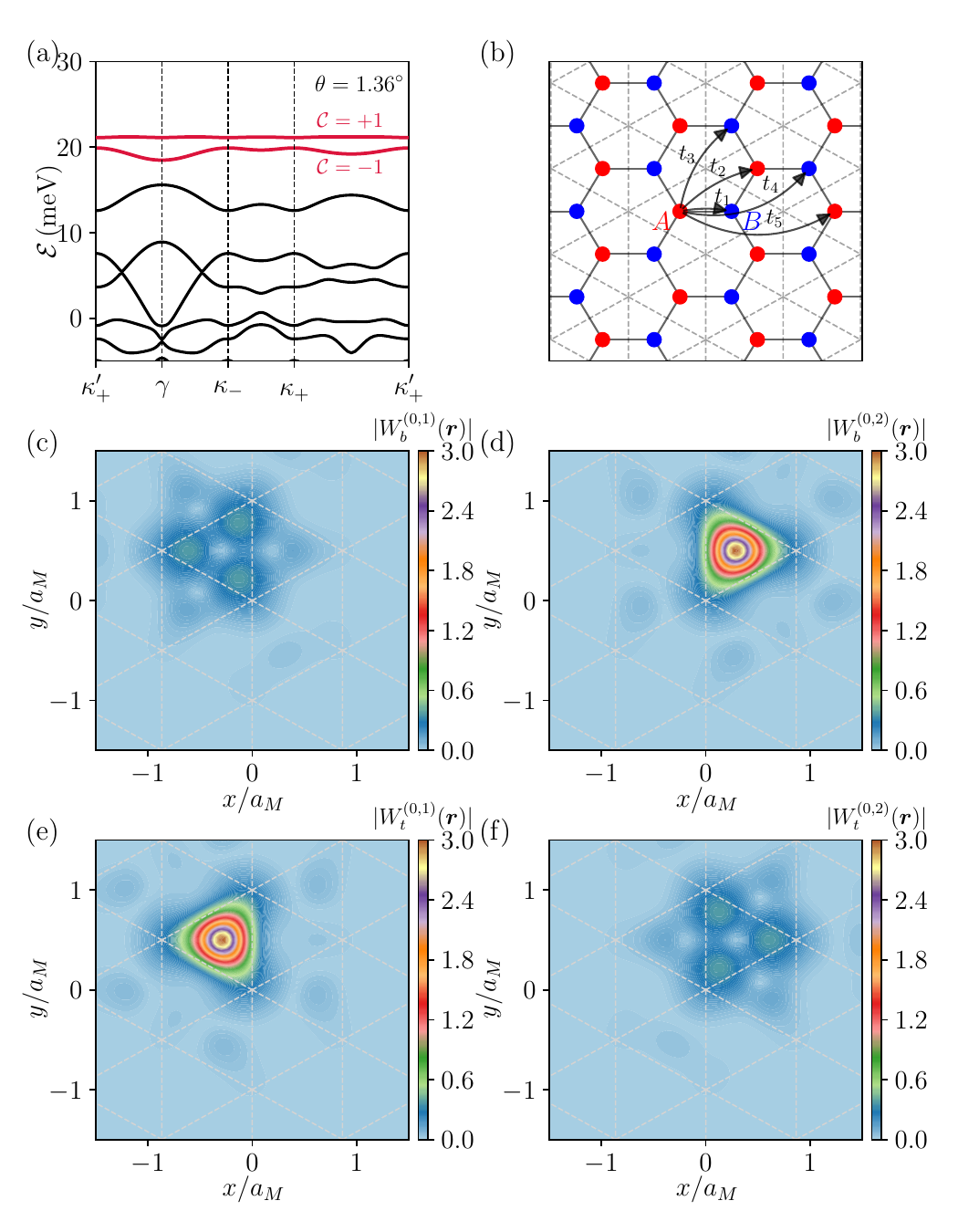}
\caption{(a) Moir\'e band structure of tMoTe$_2$ at magic angle $\theta=1.36^\circ$ along high-symmetry path in moir\'e Brillouin zone. The first and second moir\'e bands, marked in red, possess Chern numbers $\mathcal C=+1$ and $\mathcal C=-1$, respectively. (b) An effective honeycomb lattice model for the first and second moiré bands in (a). The red and blue dots denote the A and B sublattices where the Wannier functions are localized. The parameters $t_i$ with $i=1,...,5$ represent the first five hopping terms starting from A sublattice. (c-f) Maps of $|W_{l}^{(\bl R,a)}(\bl r)|$ at $\bl R=\bl 0$, for $l=b,t$ and $a=1,2$. Parameters are from Ref.~\cite{Wu2019Topological}.} 
\label{fig:tMoTe2}
\end{figure}

Here we present the quantum geometric properties of the original model, as well as the band structures and the quantum geometric properties of the truncated models. 

Figure \ref{fig:cur_trg} displays the quantum geometry of the original model for selected bands: Figs.~\ref{fig:cur_trg}(a-b) correspond to the generalized 0LL in the $N=2$ and $N=3$ models, Figs.~\ref{fig:cur_trg}(c-d) to the ideal higher Chern bands in the $N=3$ and $N=4$ models, Figs.~\ref{fig:cur_trg}(e-f) to the generalized 1LL in the $N=3$ model, and Figs.~\ref{fig:cur_trg}(g-h) to the generalized 1LL in the $N=4$ model. Due to the ideal quantum geometry of the generalized 0LL and the ideal higher Chern bands, $\mathrm{Tr}[g_{\bl k}]$ equals exactly to $|\Omega_{\bl k}|$ in Figs.~\ref{fig:cur_trg}(a-d). As shown in Figs.~\ref{fig:cur_trg}(a-b), the variation of the quantum geometry in the generalized 0LL is reduced in the $N=3$ model compared to the $N=2$ model. Similarly, as shown in Figs.~\ref{fig:cur_trg}(e-h), the variation of both $\Omega_{\bl k}$ and $\mathrm{Tr}[g_{\bl k}]$ in the generalized 1LL are smaller in the $N=4$ model than in the $N=3$ model. These results suggest that a more uniform lattice sampling of real space helps smooth the quantum geometry. 

We then focus on the truncated model, where hopping amplitudes are restricted to distances not exceeding the cutoff $d = 2a$.
The band structures of the truncated models (orange lines) for $N = 2$, $3$, and $4$ are shown in Fig.~\ref{fig:truncated_bands}, with the original models (blue lines) included for comparison; the two exhibit semi-quantitative agreement. In Figs.~\ref{fig:cur_trg_trunc1} and~\ref{fig:cur_trg_trunc2}, we further present the quantum geometry of the corresponding bands. The generalized 0LL and the ideal higher Chern band in the truncated models display nearly ideal quantum geometry. Furthermore, the quantum geometric quantities show close agreement between the truncated and original models.

\section{realization of the $N=2$ model}
In twisted bilayer MoTe$_2$ (tMoTe$_2$), the two topmost moiré valence bands can be mapped onto an effective honeycomb lattice model \cite{Wu2019Topological}. In the following, we demonstrate that this effective model at the magic angle quantitatively realizes the $N=2$ model. 
We numerically obtain the effective tight-binding model for tMoTe$_2$ by constructing the Wannier states and calculate the corresponding hopping amplitudes, which we then compare with those of the $N=2$ model.

% \subsection{Minimal two-band model of twisted MoTe$_2$}
\begin{figure*}[t]   
\includegraphics[width=1.75\columnwidth]{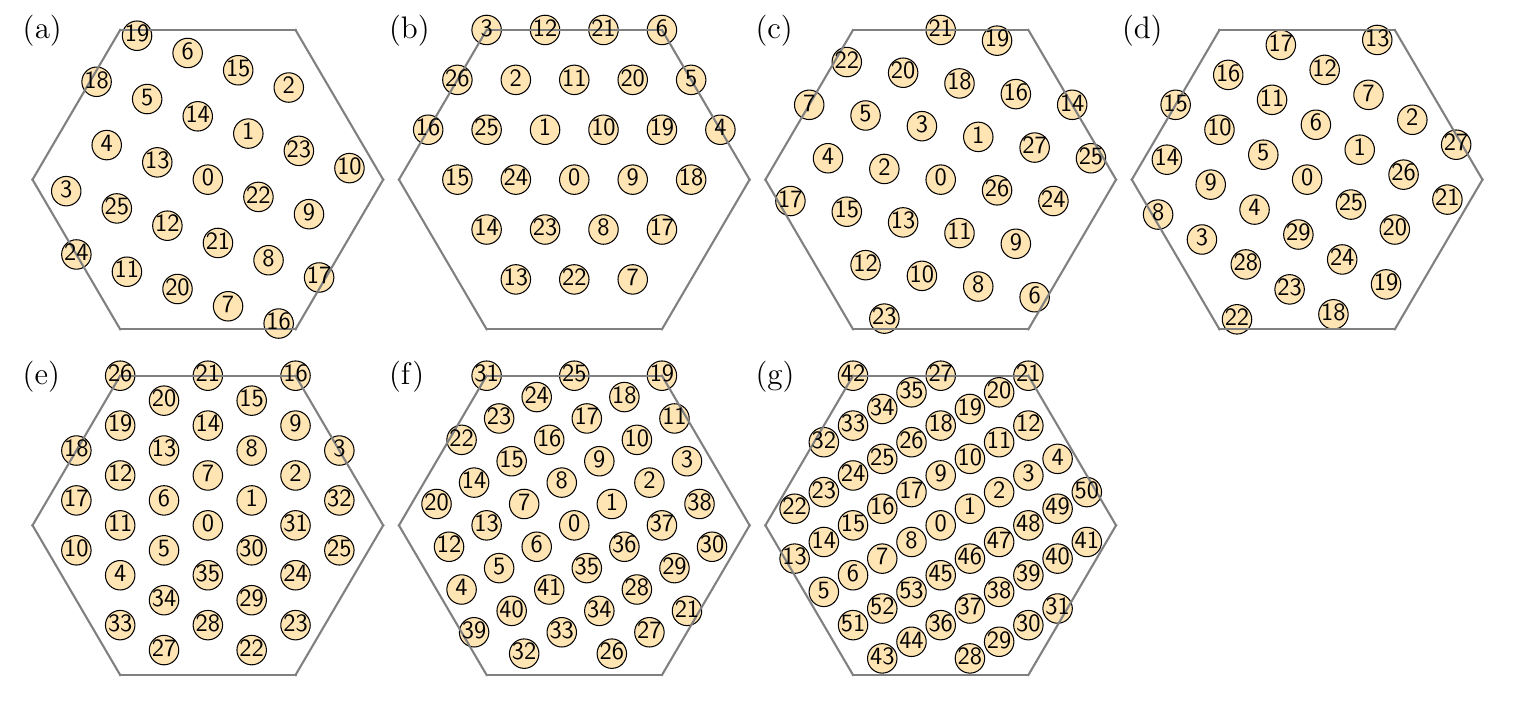}
\caption{The momentum clusters of sizes $N_s=26$, $27$, $28$, $30$, $36$, $42$, and $54$ used in the ED calculations. The circled numbers are momentum indices.} 
\label{fig:ED_grid_total}
\end{figure*}

We start from the continuum model for the valence states in tMoTe$_2$ at the $+K$ valley. The moiré Hamiltonian is given by
\begin{equation}
H_{\mathrm{cont}} = \begin{pmatrix}
-\frac{\hbar^2(\hat{\bl k}-\bm{\kappa}_+)^2}{2m^*}+\Delta_+(\bl r) & \Delta_t(\bl r) \\
\Delta_t^\dagger(\bl r) & -\frac{\hbar^2(\hat{\bl k}-\bm{\kappa}_-)^2}{2m^*}+\Delta_-(\bl r)
\end{pmatrix},
\end{equation}
where $\hat{\bl k}$ is the momentum operator, $m^*$ is the effective mass, and $\bm{\kappa}_\pm = \frac{4\pi}{3a_M}(-\frac{\sqrt{3}}{2},\mp\frac{1}{2})$ are the corners of the moiré Brillouin zone. The moiré period $a_M \approx a_0/\theta$ is determined by the twist angle $\theta$ and the monolayer lattice constant $a_0$. The layer-dependent moiré potentials $\Delta_\pm(\bl r)$ and the inter-layer tunneling $\Delta_t(\bl r)$ are
\begin{equation}
\begin{aligned}
\Delta_\pm(\bl r) &= 2V_1\sum_{j=1,3,5}\cos(\bl g_j\cdot\bl r \pm \psi), \\
\Delta_t(\bl r) &= w(1+e^{-i\bl g_2\cdot\bl r}+e^{-i\bl g_3\cdot\bl r}),
\end{aligned}
\end{equation}
with moiré reciprocal lattice vectors $\bl g_i = \frac{4\pi}{\sqrt{3}a_M}[\cos\frac{\pi(i-1)}{3},\sin\frac{\pi(i-1)}{3}]$. We adopt three sets of parameters from the literature as listed in Table.~\ref{tab:parameters_tMoTe2}. 

\begin{table}[b]
    \centering
    \caption{Parameters for the continuum model of tMoTe$_2$.}
    \label{tab:parameters_tMoTe2}
    \begin{tabular}{l|ccccc}
        \toprule
        \textbf{Parameters} & $a_0$\,(nm) & $m^*$ & $\psi$\,(deg) & $V_1$\,(meV) & $w$\,(meV) \\
        \midrule
        Ref.~\cite{Wu2019Topological} Param. & 0.3472 & 0.62\,$m_e$ & -89.6 & 8 & -8.5 \\
        Ref.~\cite{Reddy2023Fractional} Param. & 0.352 & 0.62\,$m_e$ & -91 & 11.2 & -13.3 \\
        Ref.~\cite{Wang2024Fractional} Param. & 0.352 & 0.6\,$m_e$ & -107.7 & 20.8 & -23.8 \\
        \bottomrule
    \end{tabular}
\end{table}

Figure~\ref{fig:tMoTe2}(a) shows the moir\'e band structure at the magic angle $\theta = 1.36^\circ$ corresponding to the parameter set in Ref.~\cite{Wu2019Topological}, where the magic angle is defined as the twist angle at which the bandwidth of the first moir\'e band attains its minimum and has nearly ideal quantum geometry \cite{Devakul2021Magic}. The first two moir\'e bands are separated from other bands by a finite energy gap and carry opposite Chern numbers $\mathcal C = \pm 1$. To build up an effective lattice model for these two bands, we construct localized Wannier states,
\begin{equation}
W^{(\wt{\bl R},a)}(\bl r) = \frac{1}{\sqrt{N_{\bl k}}}\sum_{\bl k}e^{-i\bl k\cdot\wt{\bl R}}\varphi_{\bl k}^{(a)}(\bl r),
\end{equation}
where $\wt{\bl R}$ denotes the unit-cell vector of the moir\'e superlattice, $N_{\bl k}$ is the number of $\bl k$ points in moir\'e Brillouin zone, $a=1,2$ labels the Wannier state centered at sites A and B,  respectively. The positions of A and B sublattices relative to $\wt{\bl R}$ are given by ${\bf r}_A = \frac{a_M}{\sqrt{3}}[-\frac{1}{2},\frac{\sqrt{3}}{2}]^\mathsf{T}$ and ${\bf r}_B = \frac{a_M}{\sqrt{3}}[\frac{1}{2},\frac{\sqrt{3}}{2}]^\mathsf{T}$.
$\varphi_{\bl k}^{(a)}(\bl r)$ is defined by
\begin{equation}
\varphi_{\bl k}^{(a)}(\bl r) = \sum_{n=1,2} \wt\psi_{\bl k,n}(\bl r) V_{\bl k,n,a},
\end{equation}
where $\wt\psi_{\bl k,n}(\bl r)$ is the Bloch wave function of tMoTe$_2$ for the $n$th moir\'e band and $V_{\bl{k}}$ is the $2\times 2$ unitary matrix used to disentangle the layer hybridization. We determine $V_{\bl{k}}$ such that $\varphi_{\bl k}^{(1)}$ ($\varphi_{\bl k}^{(2)}$) is maximally polarized to the top (bottom) layer. This maximum value problem is equivalent to finding the eigenstates of the layer polarization operator $\sigma_z$, expressed as the $z$ Pauli matrix in the layer pseudospin space, projected to the subspace spanned by $\wt\psi_{\bl k,n}$ with $n=1,2$,
\begin{equation}
\begin{aligned}
\Pi_{\bl k} = \begin{pmatrix} 
\langle\wt\psi_{\bl k,1}|\sigma_{z}|\wt\psi_{\bl k,1}\rangle & \langle\wt\psi_{\bl k,1}|\sigma_{z}|\wt\psi_{\bl k,2}\rangle \\
\langle\wt\psi_{\bl k,2}|\sigma_{z}|\wt\psi_{\bl k,1}\rangle & \langle\wt\psi_{\bl k,2}|\sigma_{z}|\wt\psi_{\bl k,2}\rangle
\end{pmatrix},
\end{aligned}
\end{equation}
where $\wt\psi_{\bl k,n} = [\wt\psi_{\bl k,n,b},\wt\psi_{\bl k,n,t}]^T$ is a two-component spinor in the layer-pseudospin space. Then, $V_{\bl{k}}$ is determined by
\begin{equation}
\begin{aligned}
V_{\bl{k}}^\dagger{\Pi}_{\bl{k}}V_{\bl{k}} = \left( \begin{matrix} {\rho }_{\bl{k}}^{1} & 0 \\  0 & {\rho }_{\bl{k}}^{2}\end{matrix}\right) ,
\end{aligned}
\end{equation}
where ${\rho }_{\bl k}^{1} < {\rho }_{\bl k}^{2}$ . We further fix the phase of $\varphi_{\bl k}^{(a)}(\bl r)$ in
the following way. The phase of $\varphi_{\bl k}^{(1)}(\bl r)$ is chosen by requiring that its top layer component is real and positive at site A position ${\bf r}_A$. Similarly, we take the bottom layer component of $\varphi_{\bl k}^{(2)}(\bl r)$ to be real and positive at site B position ${\bf r}_B$.

\begin{table}[b]
\centering
\caption{Relative hopping parameters $\wt{t}_{2\sim 5}$ in the minimal two-band model for two parameter sets, and $\wt{J}_{2\sim 5}$ in the $N=2$ model. The magic angles for the parameter sets \cite{Wu2019Topological,Reddy2023Fractional,Wang2024Fractional} are $1.36^\circ$, $1.56^\circ$ and $2.98^\circ$, respectively.}
\label{tab:parameters}
\begin{tabular}{l|cccc}
\toprule
\textbf{Parameters} & $\wt {t}_2$ & $\wt {t}_3$ & $\wt {t}_4$ & $\wt {t}_5$ \\
\midrule
Ref.~\cite{Wu2019Topological} Param. & $-0.1566 - 0.2472j$ & $-0.1636$ & $0.0332$ & $-0.0004$ \\
Ref.~\cite{Reddy2023Fractional} Param. & $-0.1581 - 0.2515j$ & $-0.1679$ & $0.0364$ & $-0.0005$ \\
Ref.~\cite{Wang2024Fractional} Param. & $-0.1733-0.2576j$ & $-0.1253$ & $0.0421$ & $-0.0072$ \\
\hline 
\\[-8pt] 
 & $\wt {J}_2$ & $\wt {J}_3$ & $\wt {J}_4$ & $\wt {J}_5$ \\
\midrule
$N=2$ model & $-0.1492 - 0.2585j$ & $-0.1630$ & $0.0266$ & $0.0079$ \\
\bottomrule
\end{tabular}
\end{table}

The layer-resolved Wannier functions $|W_{l}^{(\bl{0},a)}(\bl r)|$ for $a=1,2$ and layer index $l=b,t$ are shown in Fig.~\ref{fig:tMoTe2}(c-f). We now obtain the effective honeycomb lattice model by projecting the continuum Hamiltonian of tMoTe$_2$ onto these Wannier functions. The Hamiltonian takes the form,
\begin{equation}
\begin{aligned}
H_{\mathrm{eff}} = \sum_{\wt{\bl R}\wt{\bl R}'}\sum_{ab} t_{ab}(\wt{\bl R}-\wt{\bl R}')\, c_{\wt{\bl R},a}^{\dagger}c_{\wt{\bl R}',b},
\end{aligned}
\end{equation}
where $c_{\wt{\bl R},a}^{\dagger}$($c_{\wt{\bl R},a}$) is the creation (annihilation) operator of Wannier state and $a= 1,2$ labels the Wannier orbital. $t_{ab}(\wt{\bl R})$ is given by
\begin{equation}
\begin{aligned}
t_{ab}(\wt{\bl R}) = & \int d^2\bl r\; [W^{(\wt{\bl R},a)}(\bl r)]^*H_{\mathrm{cont}} W^{(\bl{0},b)}(\bl r) \\
= & \frac{1}{N_{\bl k}}\sum_{\bl k}\sum_{n=1,2}  e^{i\bl k\cdot\wt{\bl R}} V_{\bl k,n,a}^* V_{\bl k,n,b} \varepsilon_{\bl k,n},
\end{aligned}
\end{equation}
where $\varepsilon_{\bl k,n}$ is the energy of the $n$th moir\'e valence band of tMoTe$_2$ at $+K$ valley.

% \subsection{Comparison of hopping parameters}

%We calculate $t_{ab}(\wt{\bl R})$ for the three sets of parameters \cite{Wu2019Topological,Reddy2023Fractional,Wang2024Fractional} at the magic angles $1.36^\circ$, $1.56^\circ$ and $2.98^\circ$, respectively.
We compute the hopping amplitudes $t_{ab}(\wt{\bl R})$ using three distinct parameter sets from Refs.~\cite{Wu2019Topological,Reddy2023Fractional,Wang2024Fractional}, corresponding to the magic angles $1.36^\circ$, $1.56^\circ$, and $2.98^\circ$, respectively.
We define $\wt {\bl R}_{1,2}=a_M[\frac{\sqrt{3}}{2},\mp\frac{1}{2}]$ as the primitive lattice vectors. The first five representative hopping parameters of the effective lattice model, denoted by $t_n$ for $n=1,...,5$, are chosen as 
\begin{equation}
\small
t_{BA}(0),t_{AA}(\wt{\bl R}_{2}),t_{BA}(-\wt{\bl R}_{1}+\wt{\bl R}_{2}),t_{BA}(\wt{\bl R}_{2}),t_{AA}(\wt{\bl R}_{1}+\wt{\bl R}_{2}).
\end{equation}
Figure.~\ref{fig:tMoTe2}(b) shows the hopping path of $t_n$ for $n=1,...,5$. For the corresponding hopping parameters in the $N=2$ model, we choose $J_{n}$ for $n=1,...,5$ as
\begin{equation}
\small
J_{BA}(0),J_{AA}(\bl R_{2}),J_{BA}(-\bl R_{1}+\bl R_{2}),J_{BA}(\bl R_{2}),J_{AA}(\bl R_{1}+\bl R_{2}).
\end{equation} 
We further define the hopping parameters relative to the nearest hopping term $\wt{t}_n \equiv t_n/t_1$ and $\wt{J}_n \equiv J_n/J_1$ for both models. In table \ref{tab:parameters}, we compare $\wt{t}_n$ and $\wt{J}_n$ for $n=2,...,5$. The numerical results show a quantitative match of $\wt{t}_n$ and $\wt{J}_n$ at $n$=2, 3, and 4, while the other terms are numerically negligible. Here we choose a gauge such that $t_1$ and $J_1$ have opposite signs. We note that the topmost moir\'e band in tMoTe$_2$ has nearly ideal quantum geometry and nearly vanishing bandwidth, corresponding to the lowest-energy band in the $N=2$ model. 

\section{momentum geometry}
In Fig.~\ref{fig:ED_grid_total}, we show the momentum clusters of sizes $N_s=26$, $27$, $28$, $30$, $36$, $42$, and $54$, used in the ED calculations.

\section{integer anomalous Hall crystal}

We study the integer anomalous Hall crystal (AHC) state at $\nu=1/2$ in the ideal higher Chern band of the $N=3$ model. This state possesses an emergent SU(2) degree of freedom parameterized on a Bloch sphere. We first obtain the polarized AHC states along the X, Y, and Z directions of the Bloch sphere. To this end, we construct the ideal subbands by folding the ideal higher Chern band along three directions $\bl G_{1,2,3}$ with $\bl G_3=-\bl G_1-\bl G_2$. The polarized AHC states are defined as the product states of the filled ideal subband, featuring nontrivial topology and an enlarged $1\times 2$ periodicity. We then obtain the AHC state on the Bloch sphere by rotating the ideal subbands associated with one folding direction, featuring nontrivial topology and an enlarged $2\times 2$ periodicity. 

% We then obtain the AHC state as the product state of the filled two ideal subbands in the quarter Brillouin zone ($\mathrm{\frac{1}{4}BZ}$). These ideal subbands are constructed by folding the ideal higher Chern band along $\bl G_1$ and $\bl G_2$, featuring an enlarged $2\times 2$ periodicity.

\subsection{Polarized AHC states}

\begin{figure}[t]   
\includegraphics[width=1.\columnwidth]{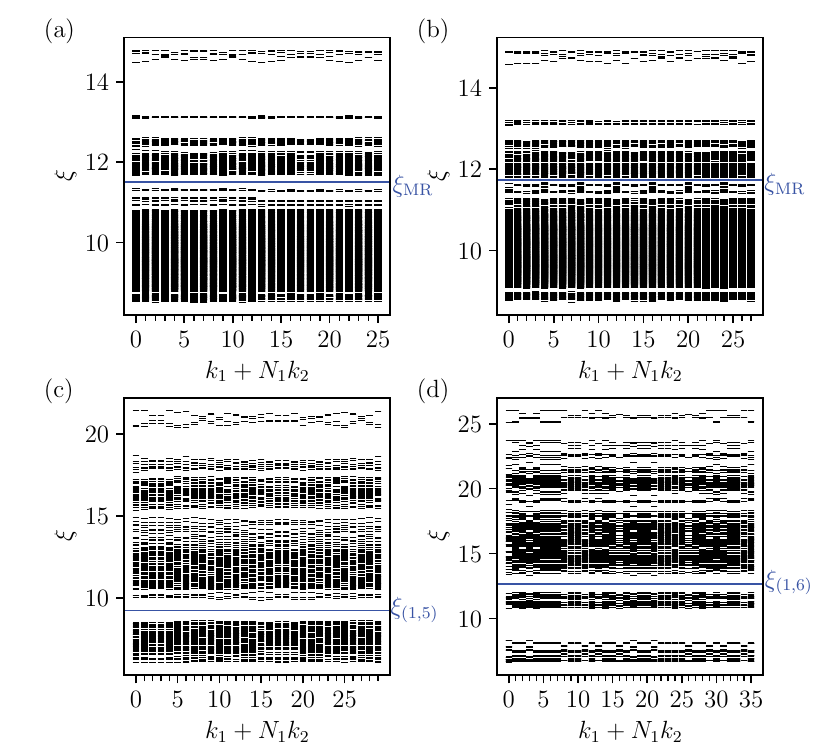}
\caption{(a-b) PES at $\nu=1/2$ in the generalized 1LL for $N=4$ model. We use $N_A = 4$ and clusters of sizes $N_s=26$ and $28$, respectively. (c-d) PES at $\nu=1/5$ and $\nu=1/6$ in the ideal higher Chern band for $N=3$ model. We use $N_A = 3$ and clusters of sizes $N_s=30$ and $36$, respectively. } 
\label{fig:PES}
\end{figure}

We derive the polarized AHC states by first constructing the subband in the half Brillouin zone (hBZ) spanned by $\bl M_n$ and $\bl G_{n+1}$, where $\bl M_n=\bl G_n/2$, $n=1,2,3$, and the index $n+1$ is understood modulo $3$. The corresponding subband Bloch state has periods $2\bl R_n$ and $\bl R_{n+1}$, where $\bl R_3=-\bl R_1-\bl R_2$. The ideal higher Chern band in the $N=3$ model can be decomposed into subbands that are related by lattice translations \cite{Wang2023Origin},
\begin{equation}
2S_{n,\bl k}\varphi_{\bl k}(\bl r)= \chi_{n,\bl k}(\bl r)+e^{-i\bl k\cdot\bl R_n}\chi_{n,\bl k}(\bl r+\bl R_n),
\label{eq:def_chi}
\end{equation}
where $\varphi_{\bl k}(\bl r)= \mathcal N_{\bl k}^{-1}\langle\bl r\ket{\Phi_{2,\bl k}}$ with $\mathcal N_{\bl k}=\mathcal N_{0,\bl k}\mathcal N_{1,\bl k}$, $\chi_{n,\bl k}(\bl r)= \langle\bl r\ket{\chi_{n,\bl k}}$ is the subband Bloch wave function and $S_{n,\bl{k}}$ is the normalization factor. Here, the Bloch state of ideal higher Chern band satisfies the quasi-periodic boundary condition in momentum space,
\begin{equation}
\Phi_{2,\bl k+\bl G_i}^\dagger=e^{-i\ell^2\bl G_i\times \bl k}\Phi_{2,\bl k}^\dagger.
\label{eq:boundary_highC}
\end{equation}
where $\Phi_{2,\bl k}^\dagger$ is the creation operator of the Bloch state.
We define $\chi_{n,\bl k}^\dagger$ as creation operator of the subband. We choose $\chi_{n,\bl k}^\dagger$ so that $\chi_{n,\bl k}^\dagger$ inherits the boundary condition along $\bl G_{n+1}$, while featuring a quasi-periodic boundary condition along $\bl M_{n}$,
\begin{equation}
\begin{aligned}
\chi_{n,\bl k+\bl M_n}^\dagger &= e^{-i\ell^2\bl M_n\times\bl k} \chi_{n,\bl k}^\dagger,\\
\chi_{n,\bl k+\bl G_{n+1}}^\dagger &= e^{-i\ell^2\bl G_{n+1}\times\bl k} \chi_{n,\bl k}^\dagger. 
\end{aligned}
\label{eq:boundary_chi}
\end{equation}
From Eqs.~\eqref{eq:def_chi}, \eqref{eq:boundary_highC}, and \eqref{eq:boundary_chi}, one solves for $\chi_{n,\bl k}^\dagger$
\begin{equation}
\begin{aligned}
\small
\chi_{n,\bl k}^\dagger 
=& S_{n,\bl k}\bigl(\varphi_{\bl k}^\dagger + e^{i\ell^2\bl M_n\times\bl k}\varphi_{\bl k+\bl M_n}^\dagger\bigr),
\end{aligned}
\label{eq:chi_dagger}
\end{equation}
where $\varphi_{\bl k}^\dagger=\mathcal N_{\bl k}^{-1}\Phi_{2,\bl k}^\dagger$ and $S_{n,\bl k}=(\mathcal N_{\bl k}^{-2} + \mathcal N_{\bl k+\bl M_n}^{-2})^{-1/2}$. 

\begin{figure}[t]
\includegraphics[width=1.\columnwidth]{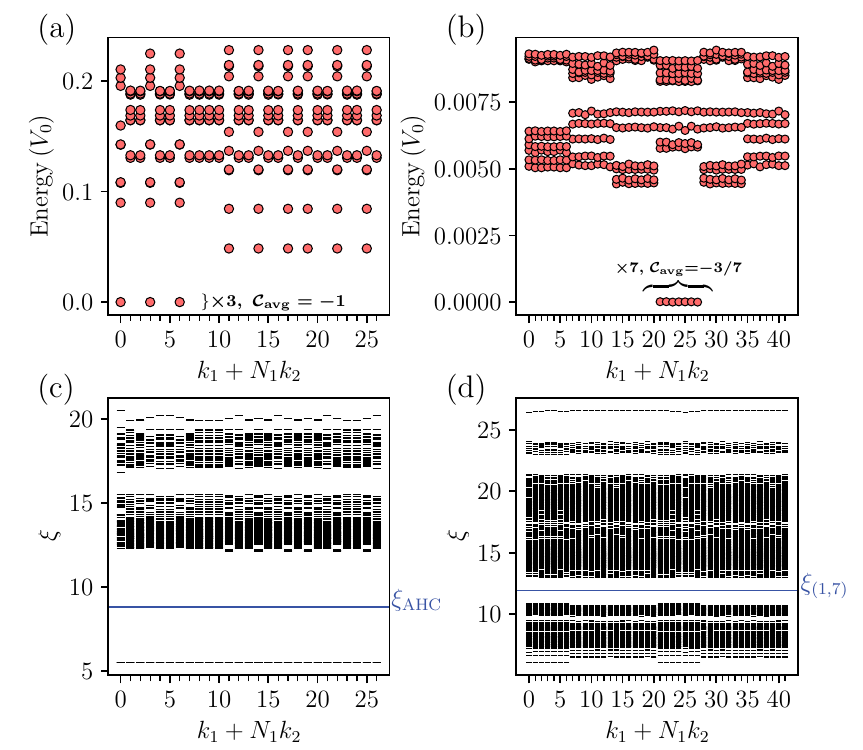}
\caption{(a-b) ED spectra at $\nu=1/3$ and $1/7$ in the $\mathcal C=-3$ band of the $N=4$ model. The system sizes are 27 and 42 unit cells. (c-d) PES with $N_A = 3$ for the corresponding quasi-degenerate states.}
\label{fig:SM_ED_ABCO}
\end{figure}

We now demonstrate that our subband construction, $\chi_{n,\bl k}^\dagger$, satisfies the ideal condition. To verify this, we show that $\chi_{n,\bl k}^\dagger$ is anti-holomorphic in momentum space up to a normalization factor $S_{n,\bl k}e^{-\frac{\ell^2}{2}|z_{\bl k}|^2}$. We note that $\varphi_{\bl k}^\dagger$ admits the decomposition,
\begin{equation}
\begin{aligned}
\small
\varphi_{\bl k}^\dagger=&\mathcal N_{\bl k}^{-1}\det [U(\bl k)]e^{-\frac{1}{2}|z_{\bl k}|^2\ell^2}\sum_sg_{s}^*(z_{\bl k})\psi_{\bl k,s}^\dagger\\
=&e^{-\frac{1}{2}|z_{\bl k}|^2\ell^2}\sum_sg_{s}^*(z_{\bl k})\psi_{\bl k,s}^\dagger,
\end{aligned}
\end{equation}
where we use Eq.~\eqref{eq:decomposition_highC} and $\det [U(\bl k)]=\mathcal N_{\bl k}$. 
As a consequence, $e^{i\ell^2\bl k'\times\bl k}\varphi_{\bl k+\bl k'}^\dagger$ is an anti-holomorphic function of $z_{\bl k}$ up to a normalization factor $e^{-\frac{\ell^2}{2}|z_{\bl k}|^2}$, 
\begin{equation}
\begin{aligned}
\small
e^{i\ell^2\bl k'\times\bl k}\varphi_{\bl k+\bl k'}^\dagger=&e^{-\frac{1}{2}|z_{\bl k}|^2\ell^2}\sum_s[\wt g_{s}(z_{\bl k}, z_{\bl M_n})]^*\psi_{\bl k,s}^\dagger\\
\wt g_{s}(z_{\bl k}, z_{\bl k'})=&e^{-\frac{1}{2}|z_{\bl k'}|^2\ell^2}e^{-z_{\bl k}z_{\bl k'}^*\ell^2}g_{s}(z_{\bl k}+ z_{\bl k'}),
\end{aligned}
\label{eq:main_decomposition}
\end{equation}
where $\wt g_{s}(z_{\bl k}, z_{\bl k'})$ is a holomorphic function of $z_{\bl k}$ and we use
\begin{equation}
\label{eq:phase_factor}
\begin{small}
\begin{aligned}
\exp(i\bl k' \times \bl k) = \exp[\frac{1}{2}|z_{\bl k}+z_{\bl k'}|^2 -\frac{1}{2}|z_{\bl k}|^2-\frac{1}{2}|z_{\bl k'}|^2-z_{\bl k'}z_{\bl k}^*].
\end{aligned}
\end{small}
\end{equation}
It then follows from Eqs.~\eqref{eq:chi_dagger} and~\eqref{eq:main_decomposition} that $\chi_{n,\bl k}^\dagger$ is anti-holomorphic in $z_{\bl k}$, up to the normalization factor $S_{n,\bl k}e^{-\frac{\ell^2}{2}|z_{\bl k}|^2}$.

% \begin{equation}
% \begin{aligned}
% \small
% \chi_{n,\bl k}^\dagger=&S_{n,\bl k}[e^{-\frac{1}{2}|z_{\bl k}|^2}\sum_sg_{s}^*(z_{\bl k})\psi_{\bl k,s}^\dagger + \\
% &e^{i\bl M_n\times\bl k}e^{-\frac{1}{2}|z_{\bl k}+z_{\bl M_n}|^2}\sum_sg_{s}^*(z_{\bl k}+z_{\bl M_n})\psi_{\bl k+\bl M_n,s}^\dagger]\\
% = &S_{n,\bl k}e^{-\frac{1}{2}|z_{\bl k}|^2}[\sum_sg_{s}^*(z_{\bl k})\psi_{\bl k,s}^\dagger+\\
% &e^{-\frac{1}{2}|z_{\bl M_n}|^2-z_{\bl k}^*z_{\bl M_n}}\sum_sg_{s}^*(z_{\bl k}+z_{\bl M_n})\psi_{\bl k+\bl M_n,s}^\dagger],
% \end{aligned}
% \end{equation}

We define $\ket{\Xi_n}$ for $n=1,2,3$ as the polarized AHC states along the X, Y, and Z directions of the Bloch sphere. $\ket{\Xi_n}$ are the product states of the filled ideal subband in hBZ,
\begin{equation}
\ket{\Xi_n} = \prod_{\bl k\in\mathrm{hBZ}}\chi_{n,\bl k}^\dagger \ket{0},
\end{equation}
where $\ket{\chi_{n,\bl k}}=\chi_{n,\bl k}^\dagger\ket{0}$. The density of $\ket{\Xi_n}$ is defined as
\begin{equation}
\rho_n(\bl r) = \sum_{\bl k\in\mathrm{hBZ}}|\chi_{n,\bl k}(\bl r)|^2.
\end{equation}
In Fig.~3 of the main text, we show the real-space map of $\rho_n(\bl r)$, where the density variation is consistent with the periods $2\bl R_n$ and $\bl R_{n+1}$. 

We compute the total weight of $\ket{\Xi_n}$ in the quasi-degenerate ground-state manifold obtained from exact diagonalization on an $N_s = 28$ cluster, defined as the sum of squared overlaps with the ground states \cite{Niu2025Quantum}. This weight exceeds 99\%, demonstrating that $\ket{\Xi_n}$ provides an accurate approximation to the ground state.

%We have calculated the weight of $\ket{\Xi_n}$ within the quasi-degenerate ground-state manifold obtained from exact diagonalization on an $N_s=28$ cluster. The weight exceeds 99\%, indicating that $\ket{\Xi_n}$ is an accurate approximation to the ground state.

\subsection{AHC states on the Bloch sphere}

We now obtain the AHC state on the Bloch sphere by constructing two subbands in the quarter Brillouin zone (qBZ) spanned by $\bl M_1$ and $\bl M_2$. The subbands are obtained by applying SU(2) rotations and proper normalization to the ideal subbands along the X direction, therefore parameterized on the Bloch sphere. The creation operators of the subbands along the X direction, $\chi^{\dagger}_{1,\bl k}$ and $\chi'^{\dagger}_{1,\bl k}$, are defined as
\begin{equation}
\begin{aligned}
\wt\chi^{\dagger}_{1,\bl k}=&S_{n,\bl k}\bigl(\varphi_{\bl k}^\dagger - e^{i\ell^2\bl M_1\times\bl k}\varphi_{\bl k+\bl M_1}^\dagger\bigr),\\
\chi'^{\dagger}_{1,\bl k} =&e^{i\ell^2\bl M_{2}\times\bl k}\wt\chi^{\dagger}_{1,\bl k+\bl M_{2}},
\end{aligned}
\label{eq:chi_prime}
\end{equation}
where $\chi'^{\dagger}_{1,\bl k}$ is orthogonal to $\chi^{\dagger}_{1,\bl k}$ and satisfies the same boundary condition. The creation operators of the subbands parameterized on the Bloch sphere, $\chi^{(\alpha,\beta)\dagger}_{\bl k}$ and $\chi'^{(\alpha,\beta)\dagger}_{\bl k}$, are defined as
\begin{equation}
\begin{aligned}
\small
\chi^{(\alpha,\beta)\dagger}_{\bl k} = & h_{\bl k}\bigl(\alpha\, S_{1,\bl k}^{-1} \chi_{1,\bl k}^\dagger + \beta\, S_{1,\bl k+\bl M_2}^{-1} \chi'^{\dagger}_{1,\bl k}\bigr),\\
= & h_{\bl k}[\alpha (\varphi_{\bl k}^\dagger+e^{i\ell^2\bl M_1\times \bl k }\varphi_{\bl k+\bl M_1}^\dagger) \\
& +\beta e^{i\ell^2\bl M_2\times \bl k }(\varphi_{\bl k+\bl M_2}^\dagger-ie^{i\ell^2\bl M_1\times\bl k }\varphi_{\bl k+\bl M_1+\bl M_2}^\dagger)]\\
\chi'^{(\alpha,\beta)\dagger}_{\bl k} = & e^{-i\ell^2\bl k\times\bl M_2}\chi^{(\alpha,\beta)\dagger}_{\bl k+\bl M_2},
\end{aligned}
\label{eq:chi_alphabeta}
\end{equation}
where $h_{\bl k}$ is the normalization factor. We can parametrize $\alpha= \cos(\theta/2)$ and $\beta= \sin(\theta/2) e^{i \phi}$, where $\theta$ and $\phi$ define, respectively, polar and azimuthal angles of a Bloch sphere.
% \begin{equation}
% h_{\bl k} = \bigl[|\alpha|^2(\mathcal N_{\bl k}^{-2}+\mathcal N_{\bl k+\bl M_1}^{-2}) + |\beta|^2(\mathcal N_{\bl k+\bl M_2}^{-2}+\mathcal N_{\bl k+\bl M_1+\bl M_2}^{-2})\bigr]^{-1/2}.
% \end{equation}
Due to Eqs.~\eqref{eq:main_decomposition} and~\eqref{eq:chi_alphabeta}, $\chi^{(\alpha,\beta)\dagger}_{\bl k}$ and $\chi'^{(\alpha,\beta)\dagger}_{\bl k}$ are anti-holomorphic in $z_{\bl k}$ up to a normalization factor $h_{\bl k}e^{-\frac{1}{2}\ell^2|z_{\bl k}|^2}$, therefore have ideal quantum geometry.
% \begin{equation}
% \begin{aligned}
% \small
% \chi_{n,\bl k}^\dagger
% = &h_{\bl k}e^{-\frac{1}{2}|z_{\bl k}|^2}[\alpha\sum_sg_{s}^*(z_{\bl k})\psi_{\bl k,s}^\dagger\\
% +&\alpha e^{-\frac{1}{2}|z_{\bl M_1}|^2-z_{\bl k}^*z_{\bl M_1}}\sum_sg_{s}^*(z_{\bl k}+z_{\bl M_1})\psi_{\bl k+\bl M_1,s}^\dagger]\\
% +&\beta e^{-\frac{1}{2}|z_{\bl M_2}|^2-z_{\bl k}^*z_{\bl M_2}}\sum_sg_{s}^*(z_{\bl k}+z_{\bl M_2})\psi_{\bl k+\bl M_2,s}^\dagger]\\
% -&i\beta e^{-\frac{1}{2}|z_{\bl M_1+\bl M_2}|^2-z_{\bl k}^*z_{\bl M_1+\bl M_2}}\sum_sg_{s}^*(z_{\bl k}+z_{\bl M_1+\bl M_2})\psi_{\bl k+\bl M_1+\bl M_2,s}^\dagger]
% \end{aligned}
% \end{equation}

The AHC state $\ket{\Xi^{(\alpha,\beta)}}$ is then defined as the product state of the filled two subbands parameterized on the Bloch sphere. $\ket{\Xi^{(\alpha,\beta)}}$ is formulated as
\begin{equation}
\ket{\Xi^{(\alpha,\beta)}} = \frac{1}{S_{\alpha,\beta}} \prod_{\bl k\in\mathrm{qBZ}} \chi^{(\alpha,\beta)\dagger}_{\bl k} \chi'^{(\alpha,\beta)\dagger}_{\bl k} \ket{0},
\label{eq:Psi_alphabeta}
\end{equation}
where $S_{\alpha,\beta}$ is the normalization factor. We note that Eq.~\eqref{eq:Psi_alphabeta} is equivalent to the Eq.~(9) in the main text up to a phase factor. $\chi_{n,\bl k}^\dagger$ can be represented by $\chi^{(\alpha,\beta)\dagger}_{\bl k}$ and $\chi'^{(\alpha,\beta)\dagger}_{\bl k}$ on the Bloch sphere along X, Y, and Z direction,
\begin{equation}
\begin{aligned}
\chi_{1,\bl k}^\dagger &= \frac{S_{1,\bl k}}{h_{\bl k}}\chi_{\bl k}^{(1,0)\dagger},\\
\chi_{2,\bl k}^\dagger &= \frac{S_{2,\bl k}}{h_{\bl k}}\Bigl[\chi_{\bl k}^{(\frac12,\frac12)\dagger} + \chi_{\bl k}^{\prime(\frac12,\frac12)\dagger}\Bigr],\\
\chi_{3,\bl k}^\dagger &= \frac{S_{3,\bl k}}{h_{\bl k}}\Bigl[\chi_{\bl k}^{(\frac12,\frac{i}{2})\dagger} - i\,\chi_{\bl k}^{\prime(\frac12,\frac{i}{2})\dagger}\Bigr].
\end{aligned}
\end{equation}
Therefore, the three polarized AHC states $\ket{\Xi_n}$ are equivalent to the $\ket{\Xi^{(\alpha,\beta)}}$ along X, Y, and Z direction on the Bloch sphere up to a phase factor,
\begin{equation}
\small
\ket{\Xi^{(1,0)}} = e^{i\phi_1}\ket{\Xi_1},
\ket{\Xi^{(\frac12,\frac12)}} = e^{i\phi_2}\ket{\Xi_2},
\ket{\Xi^{(\frac12,\frac{i}{2})}} = e^{i\phi_3}\ket{\Xi_3}.
\end{equation}

\section{Particle entanglement spectrum}

We show in detail the particle entanglement spectrum (PES) in the $\mathcal C=-2$ band for $N=3$ model and the generalized 1LL for $N=4$ model. The PES is constructed by partitioning the system into two subsystems $A$ and $B$, containing $N_A$ and $N_B$ particles, respectively \cite{Extracting2011Sterdyniak,Chen2025Robust}. The reduced density matrix $\hat{\rho}_{A}$ of subsystem $A$ is defined as $\hat{\rho}_A = \frac{1}{N_{\mathrm{GS}}} \sum_{m=1}^{N_{\mathrm{GS}}} \hat{\rho}_{m,A}$, the average over the reduced density matrices of all quasi-degenerate ground states $\ket{\Psi_m}$. Here, $N_{\mathrm{GS}}$ is the ground-state quasi-degeneracy, and $\hat{\rho}_{m,A} = \operatorname{Tr}_B(\ket{\Psi_m}\bra{\Psi_m})$. The PES levels $\{\xi_{A,n}\}$ are then obtained via the relation $\hat\rho_{A}=\sum_{n}e^{-\xi_{A,n}}\ket{\alpha_{A,n}}\bra{\alpha_{A,n}}$, where $\ket{\alpha_{A,n}}$ are the eigenvectors of $\hat\rho_A$.

\begin{figure}[t]
\includegraphics[width=1.\columnwidth]{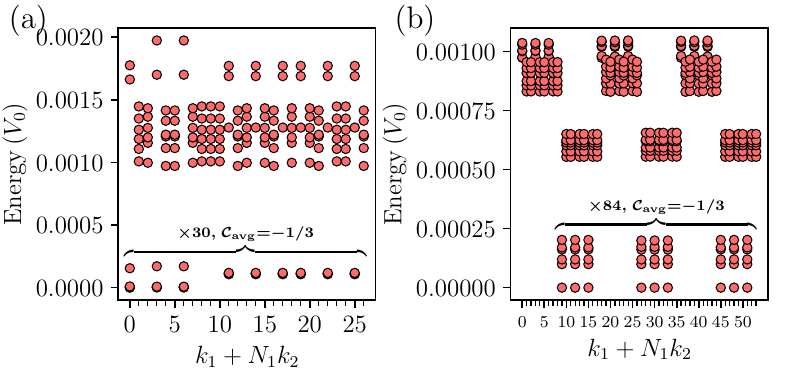}
\caption{(a-b) ED spectra at $\nu=1/9$ in the $\mathcal C=-3$ band of the $N=4$ model. The system sizes are 27 and 54 unit cells. }
\label{fig:SM_ED_ABCO2}
\end{figure}

Figures ~\ref{fig:PES}(a) and \ref{fig:PES}(b) show the PES at $\nu=1/2$ in the generalized 1LL for $N=4$ model. Here we use clusters of sizes $N_s=26$ and $28$, respectively, and choose $N_A=4$. PES gap is found in both clusters at $\xi_{\mathrm{MR}}$. The PES countings are 13338 and 18571 for the $N_s=26$ and $28$ clusters, respectively, matching the Moore-Read state.  

Figures ~\ref{fig:PES}(c) and \ref{fig:PES}(d) show the PES at $\nu=1/5$ and $\nu=1/6$ in the $\mathcal C=-2$ band for $N=3$ model. Here we use clusters of sizes $N_s=26$ and $28$, respectively, and choose $N_A=3$. For $\nu = 1/5$, a clear entanglement gap is observed at $\xi_{\mathrm{(1,5)}}$, below which the low-lying PES level count is 1360. This counting satisfies the $(1,5)$ generalized Pauli exclusion principle expected for the Halperin (332) state \cite{Wang2022Hierarchy, Dong2023Manybody, Liu2025Engineering,Liu2015NonAbelian, Liu2021Gate}. At $\nu = 1/6$, we identify a significant entanglement gap at $\xi_{\mathrm{(1,6)}}$. The counting of the low-lying PES levels below this gap is 2280, exhibiting a counting pattern that matches the generalized Pauli exclusion principle of type $(1,6)$, a characteristic feature of the Laughlin-type fractionalized state.

\section{$\mathcal C=-3$ band}
We investigate the $\mathcal C=-3$ band of the $N=4$ model by studying $\mathcal H_4$ using nearest-neighbour interaction. We first perform ED calculations at $\nu=1/3$ and $1/7$. At $\nu=1/3$, we find robust evidence of the integer anomalous Hall crystal. In Fig.~\ref{fig:SM_ED_ABCO}(a), we show the ED spectrum at $\nu=1/3$, revealing a gapped threefold quasi-degeneracy. The three nearly degenerate ground states occur at high-symmetry momenta with indices 0, 3 and 6, corresponding to $\bl{\Gamma},\bl K$ and $\bl K'$ points in the Brillouin zone, which is a characteristic signature of a charge-ordered state. The integer anomalous Hall crystal state is further confirmed by PES with $N_A=3$, as shown in Fig.~\ref{fig:SM_ED_ABCO}(b). The PES at $\nu=1/3$ exhibits a clear entanglement gap at $\xi_{\mathrm{AHC}}$, below which the low-lying level counting follows $3\binom{N_e}{N_A}=3\binom{9}{3}=252$, characteristic of a charge-ordered state. We further compute the many-body Chern number for the three quasi-degenerate ground states, yielding a quantized value $\mathcal C_{\mathrm{avg}}=-1$, consistent with the AHC phase.

At $\nu = 1/7$, the ED spectrum in Fig.~\ref{fig:SM_ED_ABCO}(b) reveals a gapped sevenfold quasi-degenerate ground-state manifold, and the PES with $N_A = 3$ displayed in Fig.~\ref{fig:SM_ED_ABCO}(d) shows a low-lying level counting of $3524$ that follows the $(1,7)$ generalized Pauli exclusion principle. These results provide evidence for a multicomponent Halperin state, which can be understood as arising from vortex attachment to the integer quantum Hall state at $\nu=1$ \cite{Dong2023Manybody}. The many-body Chern number for this state is quantized to $\mathcal{C}_{\mathrm{avg}} = -3/7$, further confirming this identification.

We then examine the $\nu=1/9$ filling in the $\mathcal C=-3$ band. Figures ~\ref{fig:SM_ED_ABCO2}(a-c) show the ED spectra using clusters of sizes $N_s=27$ and $54$. At $\nu=1/9$, the spectrum reveals a gapped $30$-fold and $84$-fold quasi-degenerate ground-state manifold for $N_s=27$ and $N_s=54$, respectively. This reflects a product structure of a threefold Laughlin-type topological degeneracy on the torus and internal multiplicity $(N_e + 2)(N_e + 1)/2$ arising from the emergent SU(3) structure of the crystal, thus realizing a fractional anomalous Hall crystal \cite{Dong2023Manybody}. The average many-body Chern number computed for both degenerate ground states take the value $\mathcal{C}_{\mathrm{avg}} = -1/3$, further confirming the fractional AHC.

In contrast to the $\nu=1/9$ case, where signatures of an emergent SU(3) structure are evident, the situation at $\nu=1/3$ appears qualitatively different. Following the construction of the integer anomalous Hall crystal in the $\mathcal{C}=-2$ band, one might expect an emergent SU(3) symmetry for the $\mathcal{C}=-3$ band at $\nu=1/3$, which would lead to a $(N_e+2)(N_e+1)/2$-fold quasi-degenerate ground-state manifold. However, our ED results at $\nu=1/3$ reveal only a threefold quasi-degenerate ground state. This suggests that the emergent SU(3) symmetry, if present, is not well developed here and may be more strongly broken compared to the lower filling $\nu=1/9$.

\bibliography{ref}